\documentclass[twocolumn]{aastex63}
\usepackage[T1]{fontenc}

\usepackage{graphicx}	
\graphicspath{{figs/}}
\usepackage{amsmath}	
\usepackage{amssymb}	

\usepackage{mhchem}
\usepackage{lipsum}
\usepackage{natbib}
\usepackage{mathtools}
\usepackage{tikz}
\usepackage{environ}
\usepackage[UKenglish]{datetime}

\usepackage{siunitx}
\usepackage{makecell}
\usepackage{breqn}
\usepackage{physics}
\usepackage[textsize=tiny]{todonotes}
\usetikzlibrary{arrows,shapes,angles,quotes}

\newcommand{\re}{R$_{\oplus}$}
\newcommand{\me}{M$_{\oplus}$}
\shorttitle{Sub-Neptune atmospheric dynamics}
\shortauthors{Innes and Pierrehumbert}
\begin{document}
\title{Atmospheric dynamics of temperate sub-Neptunes. Part I: dry dynamics}
\author[0000-0001-5271-0635]{Hamish Innes}
\affiliation{Atmospheric, Oceanic and Planetary Physics, University of
  Oxford}
\author[0000-0002-5887-1197]{Raymond T. Pierrehumbert}
\affiliation{Atmospheric, Oceanic and Planetary Physics, University
  of Oxford}

\begin{abstract}
Sub-Neptunes (planets with radii between 2 and 4 \re) are abundant
around M-dwarf stars, yet the atmospheric dynamics of these planets is
relatively unexplored. In this paper, we aim to provide a basic underpinning of
the dry dynamics of general low mean molecular weight, temperate
sub-Neptune atmospheres. We use the ExoFMS GCM with an idealised grey
gas radiation scheme to simulate planetary atmospheres with different
levels of instellation and rotation rates, using the atmosphere of
K2-18b as our control. We find that the atmospheres of tidally-locked,
temperate sub-Neptunes
have weak horizontal temperature gradients owing to their slow
rotation rates and hydrogen-dominated composition. The zonal wind
structure is dominated by high-latitude cyclostrophic jets driven by
the conservation of angular momentum. At low pressures we observe
superrotating equatorial jets, which we propose are driven by a
Rossby-Kelvin instability similar to the type seen in simulations of
idealised atmospheres with axisymmetric forcing. By viewing the flow
in tidally-locked coordinates, we find the predominant overturning
circulation to be between the dayside and nightside, and we derive
scaling relations linking the tidally-locked streamfunction and
vertical velocities to instellation. Comparing our results to the only other GCM study of K2-18b, we find significant qualitative differences in dynamics, highlighting the need for further collaboration and investigation into effects of different dynamical cores and physical parameterizations. This paper provides a baseline for studying the dry dynamics of temperate sub-Neptunes, which will be built on in part II with the introduction of moist effects.
\end{abstract}
\section{Introduction}
\label{sec:intro}
\subsection{Sub-Neptunes}
In the last decade there has been an explosion in the number of
exoplanets discovered with no solar-system analogue in terms of
radius. Before the \textit{Kepler} mission, 86\% of confirmed exoplanets had
masses larger than Neptune whereas 85\% of \textit{Kepler}-detected
planets had radii smaller than that of Neptune \citep{Batalha2014}. Exoplanets which fall
between the Earth's radius (1 \re) and Neptune's ($\approx$ 4 \re) are
generally classified as either super-Earths or sub-Neptunes depending
on whether their radii lie below or above the radius valley at 1.5 - 2.0 \re
\citep{Fulton2017} respectively. Of the 3359 discovered exoplanets
with confirmed radii, 1290 (38\%) have radius between 2 and 4 \re
(sub-Neptunes) and 1141 (34\%) have radius between 1 and 2 \re
(super-Earths), making them by far the most abundant subclasses of
exoplanet detected\footnote{Data from the Nasa Exoplanet Archive,
  \url{https://exoplanetarchive.ipac.caltech.edu}, accessed
  \formatdate{14}{5}{2021}}. Subsequent analysis of planetary
population statistics suggest that between 30 and 50\% of all main
sequence stars host at least one planet within 1-4\re
\citep{Mayor2011,Marcy2014b,Petigura2013,Winn2015}.

Whereas super-Earths are thought to be
either bare rocks or have secondary
atmospheres derived from non-\ce{H2} solids \citep{Kite2020a}, sub-Neptunes are characterised by
thick, primary atmospheres with a likely high proportion of nebular \ce{H} and
\ce{He}. The paucity of planets with intermediate radii is proposed to
be either due to the loss of \ce{H}/\ce{He} envelopes through
photo-evaporation \citep{Lopez2013,Owen2013} or core-driven
atmospheric loss \citep{Ginzburg2016,Gupta2019}.

To date only five sub-Neptune sized planets have had their
atmospheres characterised by spectroscopy: GJ-1214b
\citep{Kreidberg2014}, HD 97658b \citep{Knutson2014}, K2-18b
\citep{Benneke2019}, GJ-3470b \citep{Benneke2019b} and HD 3167c \citep{Mikal-Evans2020}. Determining
the composition of a sub-Neptune's atmosphere without spectroscopic
measurements is extremely difficult since often several different
atmospheric and core compositions can explain the same mass and radius
measurements \citep{Lopez2013, Rogers2010, Fortney2013}. For example, early work on GJ-1214b \citep{Rogers2010} demonstrated that
three different scenarios for atmospheric origin (in-situ accretion of
H/He, sublimated ices and an out-gassed atmosphere) could explain the
mass-radius data despite leading to widely varying
compositions. When spectroscopic analysis was eventually carried out \citep{Bean2011,Kreidberg2014}, the relatively flat spectrum was
attributed to the presence of clouds, which has thwarted further
attempts to constrain its constituents. In addition, recent work has
shown the importance of interactions between the atmosphere and magma
oceans in setting the metallicity of the atmosphere
\citep{Kite2020}. The redox state of the magma ocean depends on whether
the planet was formed inside or outside the ice line, which then
greatly affects the partition of volatiles between the atmosphere and core.

The measurement of K2-18b's spectrum \citep{Benneke2019, Tsiaras2019} was the first
such observation for a sub-Neptune with terrestrial-like instellation
(estimated at \SI{1368}{Wm^{-2}}). Observations from the \textit{Hubble Space
Telescope}'s Wide Field Camera 3 showed a strong
absorption feature at \SI{1.4}{\micro m}, which along with retrieved
temperature-pressure profiles was put forward as evidence of liquid
water. Several studies since have tried to constrain the composition
of the atmosphere using self consistent 1D models which directly
simulate radiative transfer, convective processes and chemistry
\citep{Madhusudhan2020,Scheucher2020,Piette2020,Blain2021,Bezard2020}. In
\cite{Madhusudhan2020}, a range of planetary compositions for K2-18b were proposed
which would fit both the observed mass-radius and spectral data. These
included rocky worlds with a significant Fe/Si core, sub-Neptunes
(with a larger icy component) and water worlds, which are majority-\ce{H2O} by composition. There were a range of valid solutions which
claimed a liquid water surface was possible -- however this has been
disputed by \cite{Scheucher2020}, who argued that the small scale
height resulting from evaporation of high mean molecular weight water
from a surface ocean would flatten spectral features and be
inconsistent with observations. In addition, several water world
solutions require a significant proportion of the core mass to be
water ($\gtrsim 80\%$ by mass), which may be unlikely from formation
considerations \citep{Nettelmann2011}. The detection of water is also
somewhat disputed, with two recent papers claiming the data are better
fit by \ce{CH4} absorption \citep{Bezard2020,Blain2021}. Despite significant variations in models,
most agree that K2-18b's outer atmosphere is at least 80\% \ce{H2}-\ce{He}
by volume and at most 15\% \ce{H2O}.

\subsection{General Circulation Model Simulations of Sub-Neptunes}

Although there have been many general circulation model (GCM) simulations of hot-Jupiters
\citep[e.g.][]{Showman2002,Rauscher2010,Perez-Becker2013,Dobbs-Dixon2013,Rauscher2014,Showman2015,Kataria2016,Komacek2016,Mayne2017,Mendonca2020}
and terrestrial exoplanets
\citep[e.g.][]{Yang2014,Kaspi2015,Koll2016a,Haqq-Misra2018,Komacek2019b,Hammond2020,Sergeev2020},
there have been comparatively fewer studies of sub-Neptune
atmospheres, with most focused on GJ-1214b. The works of
\cite{Menou2012}, \cite{Zhang2017} and \cite{Drummond2018}
investigated the effect of varying metallicity on the atmosphere of
GJ-1214b, and generally agreed in finding that increased metallicity
decreases equatorial jet width and increases the day-night temperature
contrast. The effect of clouds on the dynamics and spectra of GJ-1214
b has also been probed \citep{Charnay2015a,Charnay2015b}. The atmosphere of GJ-1214b has also been used to test the
differences between standard primitive equation GCM models and models
which integrate the equations of motion without the traditional,
hydrostatic, and shallow approximations \citep{Mayne2019}. For hot
sub-Neptunes it was found that including these non-primitive terms
could change the structure of the deep atmosphere and
also affect the location of a planet's hot spot. The deep atmosphere
on sub-Neptunes also affects the equilibration times of GCMs
\citep{Wang2020}, with equilibration times of $10^4$ to $10^5$ Earth
days often required to reach a steady state.

To date, only one published work has studied the atmosphere of K2-18b
with a GCM \citep{Charnay2021}. The circulation was found to be
dominated by an overturning day-night circulation, which controlled
the location of clouds. The cooling required to induce condensation
and cloud cover occurred either at the terminators (from radiative
cooling) or at the substellar point (from adiabatic cooling on
ascent). Which mechanism dominated depended strongly on the size and
density of cloud condensation nuclei, as well as the assumed
metallicity of the atmosphere.

Although few sub-Neptunes with hydrogen-dominated atmospheres on the
threshold of \ce{H2O}
condensation have been discovered and characterised, they may prove to
be common given the abundance of sub-Neptunes and their predicted
compositions \citep{Kite2020a}. This work aims to build an
understanding of the dynamical meteorology of such objects. The effect of condensible water vapour on
these atmospheres is another area of interest, due to water having a
higher mean molecular weight than the background \ce{H2}-He gas, in
contrast to Earth. The consequence of this on convective
adjustment has been highlighted \citep{Leconte2017}, but the effect is
also likely to be important for global dynamics. We will investigate
the effect of latent heating and compositional gradients in
sub-Neptune atmospheres in a later paper (part II). This paper establishes a baseline by exploring the dry dynamics of temperate sub-Neptunes.

We describe the model used in our experiments in section~\ref{sec:model}, detail the results in section~\ref{sec:results} and discuss their implications in section~\ref{sec:discussion}. 

\section{Model}\label{sec:model}
\subsection{ExoFMS}

The GCM used for this study is ExoFMS, which is based on GFDL's
Flexible Modelling System \citep{Lin1997,Lin2004} which uses a finite
volume dynamical core on a cubed-sphere grid with physics modules
adapted for exoplanetary study. ExoFMS has been used previously to
study the atmospheres of terrestrial exoplanets
\citep{Pierrehumbert2016,Hammond2017,Hammond2018,Pierrehumbert2019}
and gas giants \citep{Lee2020,Lee2021}.

\begin{deluxetable}{ll}
  \tablehead{\colhead{Symbol} & \colhead{Meaning}}
  \tablecaption{Symbol definitions \label{tab:symbols}}
  \startdata
  $\Phi$ & Geopotential \\
  $\boldsymbol{u}_h$  & Horizontal wind vector $(u,v)$\\
  $\grad_h\cdot$ & Horizontal divergence operator\\
  $\omega$ & Pressure velocity\\
  $T$    & Temperature\\
  $c_p$  & Heat capacity at constant pressure\\
  $R$    & Gas constant\\
  $a$    & Planetary radius\\
  $f$   & Coriolis parameter $=2\Omega \sin\phi$\\
  $\theta$ & Potential temperature $=T(p/p_0)^{R/c_p}$\\
  $\dot{Q}$ & Heating rate per unit mass  
  \enddata
\end{deluxetable}

The dynamical core solves the primitive equations with a
$\sigma\text{-}p$
vertical coordinate which follows contours of constant geopotential on
the bottom boundary and transitions into a pure pressure coordinate in
the upper atmosphere. In pressure coordinates $(t,\lambda, \phi, p)$, the primitive
equations are:
\begin{subequations}
  \label{eq:prim}
\begin{align}
  \frac{\text{D}u}{\text{D}t} - \frac{uv\tan\phi}{a} - fv &=
  -\frac{1}{a\cos\phi}\pdv{\Phi}{\lambda}, \label{eq:xmom} \\
  \frac{\text{D}v}{\text{D}t} + \frac{u^2\tan\phi}{a} + fu &=
                                                               -\frac{1}{a}\pdv{\Phi}{\phi}, \label{eq:ymom}\\
  \frac{\text{D}\theta}{\text{D}t} &= \frac{\theta}{c_pT}\dot{Q},\label{eq:therm}\\
  \pdv{\Phi}{\ln p} &= -RT, \label{eq:hydro}\\
  \grad_h\cdot\boldsymbol{u}_h + \pdv{\omega}{p} &= 0. \label{eq:cont}
\end{align}
\end{subequations}
where the symbols are defined in Table~\ref{tab:symbols}. Equations \ref{eq:xmom} and \ref{eq:ymom} represent zonal and
meridional momentum balance; \ref{eq:therm} the first law of
thermodynamics; \ref{eq:hydro} hydrostatic balance; and \ref{eq:cont}
the continuity of mass.

The diabatic heating rate $\dot{Q}$ is largely determined by the
radiative transfer scheme. We use a double-grey scheme, where the
radiation is split into two bands: shortwave (SW) and
longwave (LW). The incoming stellar radiation is assumed to be only
in the SW band and we assume the atmosphere only emits in the LW part
of the spectrum. The LW and SW opacities ($\kappa_{\text{LW}}$ and
$\kappa_{\text{SW}}$ respectively) define optical depths
($\tau_{\text{LW}},\tau_{SW}$) via the relation:
\begin{equation}
  \label{eq:opacities}
  \tau_i = \frac{2\kappa_i}{g}\frac{p}{p_0} \quad i\in(\text{SW},\text{LW})
\end{equation}
using the hemi-isotropic closure to approximate the angular dependence
of $\tau$ \citep{Pierrehumbert2010}. Here we have neglected any
pressure dependence of the two opacities, which is likely to be a
false assumption at higher pressures where collisional-induced absorption due to \ce{H2} becomes important (and scales as $\rho^2\sim p^2$ for an
isothermal atmosphere). We justify this simplifcation in section~\ref{subsec:experiment-setup}. The downwards SW flux,
$S_-$, and the upwards and downwards LW fluxes, $(F_+,F_-)$, are then calculated 
via:
\begin{align}
  \label{eq:1}
  S_- &= S_0e^{-\tau_{\text{SW}}}\\
  F_- &= \int_{0}^{\tau}\sigma
        T^4(\tau')e^{-(\tau-\tau')}\dd{\tau'}\\
  \begin{split}
  F_+ &= F_{\text{int}}e^{-(\tau_{\infty}-\tau)} \\
        &\quad +\int_{\tau}^{\tau_{\infty}}\sigma
        T^4(\tau')e^{-(\tau'-\tau)}\dd{\tau'}
      \end{split}
\end{align}
where $S_0$ is the instellation, $F_{\text{int}}$ is the longwave flux radiating upwards from the
interior of the planet, and $\sigma$ is the Stefan-Boltzmann
constant. At the lower boundary, the internal flux is added as
a temperature tendency to the lowest model layer. The upwards flux at
the lower boundary is specified as $\sigma T_b^4$, where $T_b$ is the
temperature of the lowest model layer. Since our SW opacities
correspond to bottom-boundary optical depths of much greater than
unity, we neglect the small fraction of SW flux that penetrates
the bottom boundary ($\sim$ 1 ppm). We note that the incoming SW radiation does have a zenith
angle dependence, however we are neglecting this dependence on the
grounds that the direct beam is largely converted to diffuse radiation
by scattering in the optically thick section of the atmosphere (this assumption breaks down in the upper atmosphere where the direct beam is not yet scattered). The radiative heating rate is calculated from
these fluxes as:
\begin{equation}
  \label{eq:radq}
  \dot{Q}_{\text{rad}} = g\dv{p}(F_+-F_--S_-)
\end{equation}
Enthalpy-conserving dry convective adjustment is performed if the vertical temperature
gradient $\partial_pT$ exceeds that of the dry adiabat $R/c_p$.

Unless otherwise stated, the model was run with a constant value of
the atmospheric mean molecular weight, $\mu$, which sets the value of
the gas constant $R$ through $R=R_*/\mu$, where $R_*$ is the universal
gas constant, \SI{8.314}{J K^{-1}mol^{-1}}. The heat capacity at
constant pressure, $c_p$, was then set by assuming the atmosphere
behaved as an ideal diatomic gas such that $R/c_p = 2/7$. 

\subsection{Experiment Setup} \label{subsec:experiment-setup}

We use the temperate sub-Neptune K2-18b as our control
experiment. K2-18b is the first sub-Neptune with Earth-like
instellation to have its atmosphere characterised by transmission
spectroscopy. Its radius (2.61 \re \citep{Cloutier2019}) and mass
($(8.63\pm 1.35)$\me \citep{Cloutier2019}) place it close to the peak
of the sub-Neptune population density distributions \citep{Owen2017}
making it an ideal candidate for generalised GCM experiments. 

\begin{deluxetable*}{lll}
\tablehead{\colhead{Parameter} & \colhead{Value} & \colhead{Comment}}
\tablecaption{Control experiment parameters \label{tab:control}}
\startdata
\multicolumn{3}{c}{Planetary parameters}\\
\hline
$a$ (\re) & $2.61$ & \cite{Benneke2019} \\
Rotation period (Earth days) & 32.9396 & \cite{Cloutier2017}\\
$S_0$ (\si{W m^{-2}}) & 1368 & \cite{Benneke2019}\\
$g$ (\si{m s^{-2}}) & 12.43 & \cite{Benneke2019} \\
\hline
\multicolumn{3}{c}{Atmospheric parameters}\\
\hline
$\mu$ (\si{g mol^{-1}})& 2.2 & \cite{Menou2012}, solar metallicity\\
$R$ (\si{J kg^{-1}K^{-1}})& 3779 & \cite{Menou2012}, solar metallicity\\
$R/c_p$ & $2/7$ & Ideal diatomic gas \\
$c_p$ (\si{J kg^{-1}K^{-1}}) & $R\times\frac{1}{R/c_p}$ & \nodata \\
$\kappa_{\text{LW}}$ (\si{cm^2 g^{-1}}) & \num{2e-2} & \cite{Menou2012}, solar metallicity\\
$\kappa_{\text{SW}}$ (\si{cm^2 g^{-1}}) & \num{8e-4} &
\cite{Menou2012}, solar metallicity \\
$T_{\text{int}}$ (\si{K}) & 70&
\makecell[l]{Consistent with \cite{Blain2021} \\ and
  \cite{Piette2020}}\\
\hline
\multicolumn{3}{c}{Numerical parameters}\\
\hline
Horizontal grid resolution & C48 & \makecell[l]{Each side of cubed sphere has
  \\$48\times 48$ resolution \citep{Putman2007}} \\
Vertical grid resolution & 50 & \nodata \\
Bottom boundary reference pressure (\si{Pa}) & \num{1e6} & \nodata \\
Top boundary pressure (\si{Pa}) & 10 & \nodata \\
Experiment run time (Earth days) & 20000 & \nodata \\
Substellar longitude & $0^{\circ}$ & \nodata \\
Dynamical timestep (s) & 60 -- 120  & Varied according to numerical stability
\enddata
\end{deluxetable*}

\begin{figure}
  \plotone{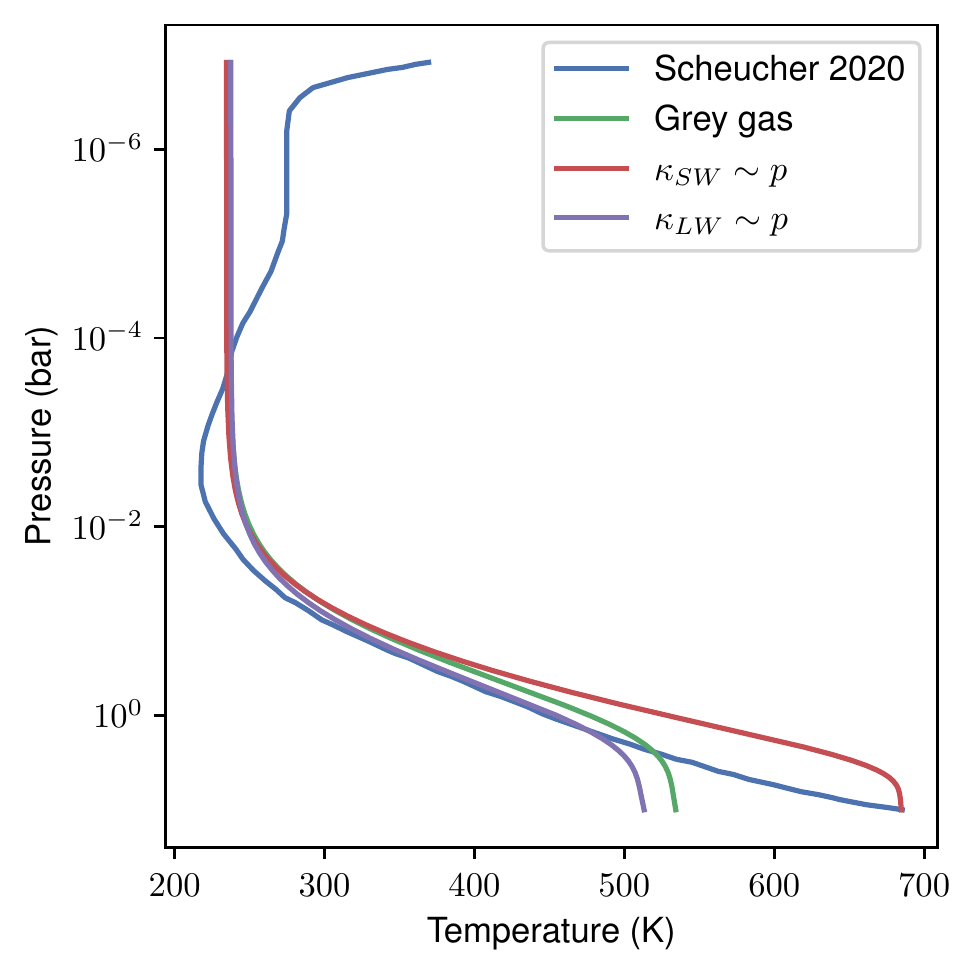}
  \caption{Comparison of the radiative equilibrium profile calculated
    using a simple grey gas scheme with a correlated-$k$ chemical
    equilibrium model from
    \cite{Scheucher2020}. We also include profiles
  calculated with pressure-dependent opacities in the LW (purple
  curve) and SW (red curve) bands. \label{fig:ggcomp}}
\end{figure}

Table~\ref{tab:control} describes the parameters for the control
run. The two-stream opacity values and mean molecular weight value
$\mu$ are taken from a study of the atmosphere of GJ-1214b
\citep{Menou2012}. A reasonable fit to the temperature-pressure
profile from \cite{Scheucher2020} was found using a column model based
on the same radiative transfer scheme (see Figure~\ref{fig:ggcomp},
green curve). The
largest discrepancy comes in the deep atmosphere,
where the grey gas solution significantly underestimates the
temperature. This is likely due to two main factors. Firstly, as
previously mentioned, collisional-induced absorption can increase the
LW opacity at the bottom of the atmosphere greatly. However, including
an additional opacity proportional to $p$ only increases the
high-pressure temperature either if there is a significant internal
flux, or if shortwave heating penetrates to the high-pressure
region. Estimates of the internal temperature of K2-18b correspond to
internal fluxes of around \SI{1}{Wm^{-2}} which is not enough to
increase the lower-layer temperatures significantly.
\medskip

We investigated the effect of adding a term proportional to
  $p$ in both LW and SW opacities such that the optical depths were of the
  form:

  \begin{equation}
    \label{eq:12}
    \tau = \tau_0\qty(f\qty(\frac{p}{p_0})^2 + (1-f)\frac{p}{p_0})
  \end{equation}
where $f$ is a constant between 0 and 1. When $f=0.8$ for the
  LW band (note we cannot set $f=1$ for the LW band since this causes
  diverging temperatures at $p=0$), we do not see a significant change in the temperature
  profiles (see purple curve in Figure~\ref{fig:ggcomp}) and the curve
  is a worse fit to the \cite{Scheucher2020} profile. When $f=1$
  for the SW band, we find the temperature is much hotter at the
  bottom of the atmosphere, better matching the \cite{Scheucher2020}
  profile at the bottom boundary. In Appendix~\ref{app:a}, we show the result of running one
  of our 6 day period experiments with this altered SW opacity. This experiment
  produces qualitatively similar results to our constant-$\kappa$
  approach, though we note the magnitude of the zonal winds when using
  pressure-dependent opacities differs notably.

We found that
including a second SW band with a lower optical depth ($\approx 1$)
was much more effective at heating the lower atmosphere. An experiment
run with this modification was found to not affect the dynamics in the
an appreciable way. 

We study the effect of varying rotation rate by running the experiment
with a faster rotation rate with $P = 6$ days. K2-18b is a
slow-rotator and lies within the ``weak temperature gradient'' regime
of parameter space \citep{Pierrehumbert2019}. Therefore while we
expect decreasing the rotation rate to not change the dynamical regime
of the atmosphere, increasing the rotation rate could produce
interesting changes to the dynamics. We choose $P=6$ days since this
is the rotation rate of TRAPPIST-1e, a planet with a temperate
equilibrium temperature of \SI{251}{K} \citep{Gillon2016} orbiting an
ultra-cool M-star with effective temperature \SI{2550}{K}
\citep{Gillon2016}. Most observed M-dwarfs have an effective
temperature greater than this value \citep{Casagrande2008,Rajpurohit2013}, so we would expect temperate
exoplanets around the majority of M-dwarfs to have longer rotation
periods than TRAPPIST-1e, making it approximately an upper limit on
rotation rate.

We also investigate changing the value of the instellation, running
the simulation with $S\in \{\frac{1}{2}S_0, S_0, 2S_0\}$ where $S_0$ is
K2-18b's instellation.

Table~\ref{tab:exp} summarises the experiments.

\begin{deluxetable}{ll}
\tablehead{\colhead{Experiment name} & \colhead{Difference to control}}
\tablecaption{Experiment setup \label{tab:exp}}
\startdata
PKc & Control experiment (see Table~\ref{tab:control})\\
PKh & $S = \frac{1}{2}S_0$ \\
PKd & $S = 2S_0$ \\
P6c & $P = 6$ days\\
P6h & $P = 6$ days, $S=\frac{1}{2}S_0$\\
P6d & $P = 6$ days, $S=2S_0$\\
\enddata
\end{deluxetable}

\section{Results} \label{sec:results}
\subsection{Convergence}
\begin{figure*}
  \gridline{\fig{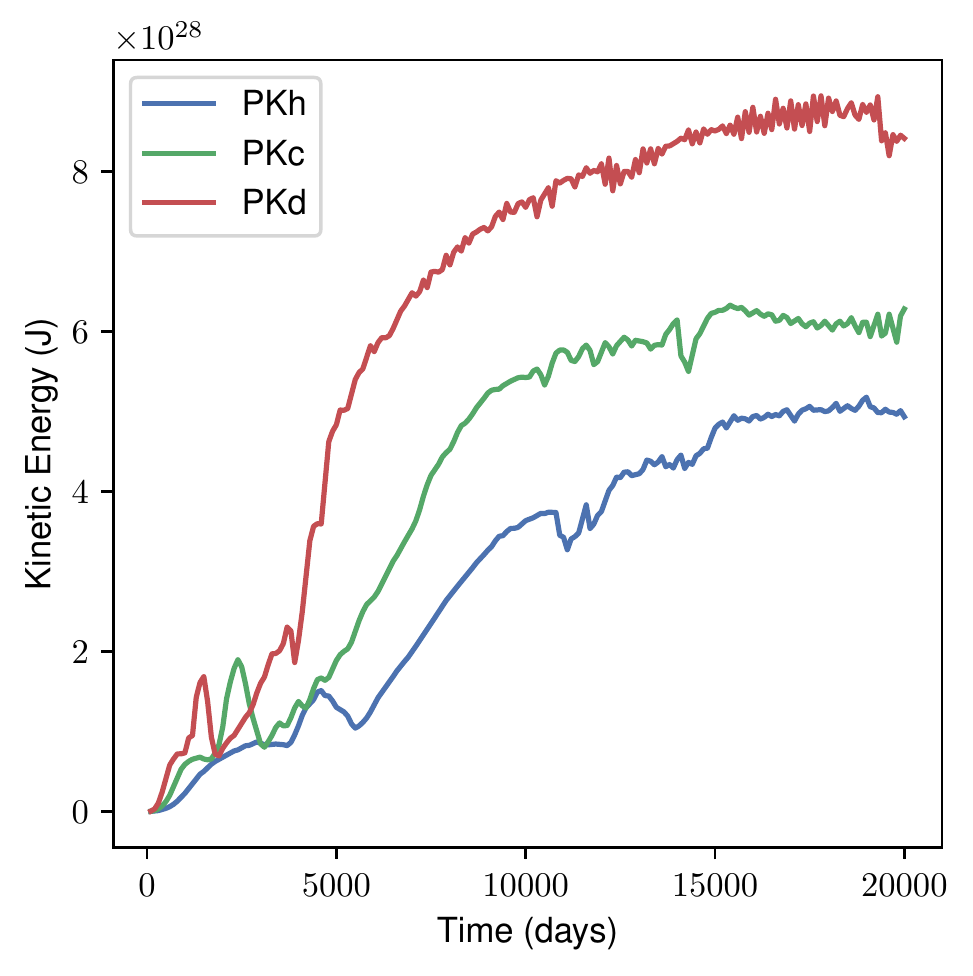}{0.5\textwidth}{(a)}
    \fig{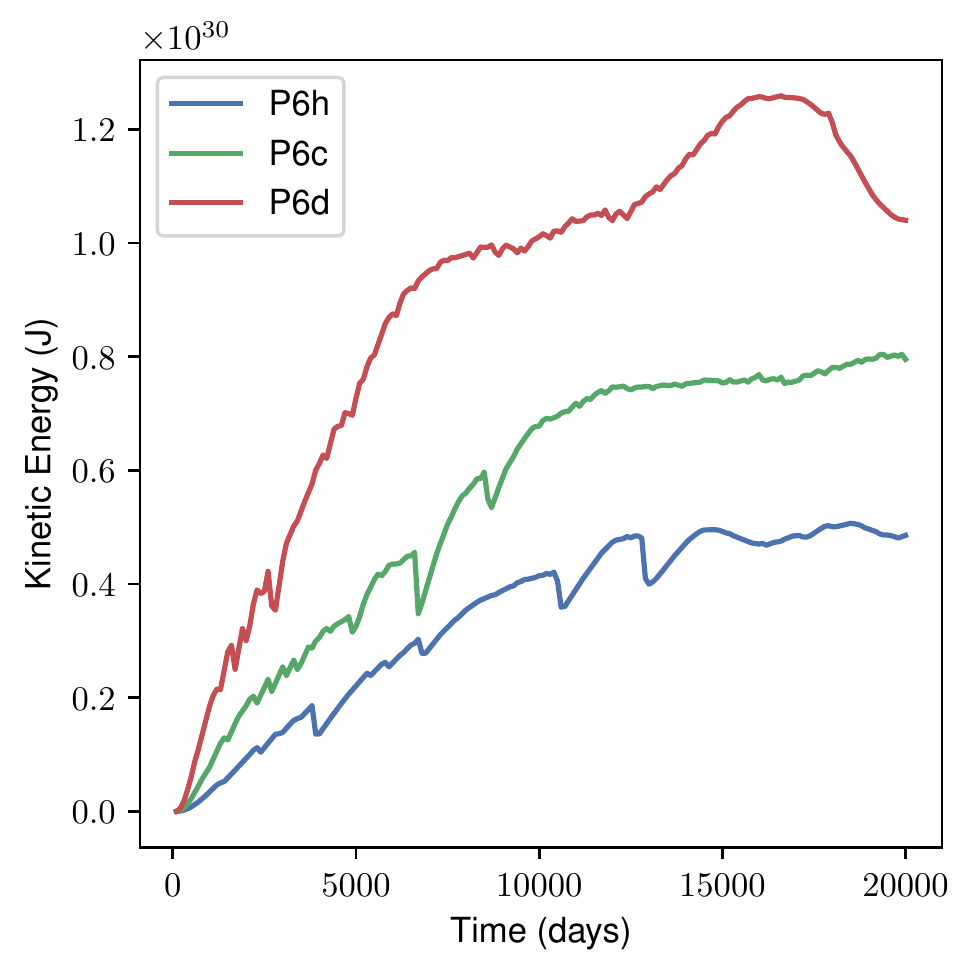}{0.5\textwidth}{(b)}}
  \caption{The kinetic energy of the dry model runs as a function of
    model time. (a) The 33 day period runs, with lines indicating
   different levels of instellation. (b) Equivalent to (a) but with
    6 day period runs. \label{fig:conv}}
  \end{figure*}

Firstly, we investigate the convergence of the model
runs. Figure~\ref{fig:conv} shows the total kinetic energy of each
model run as a function of model day, defined as:
\begin{equation}
  \label{eq:ke}
  E_k = \int_A\int_0^{p_s}\frac{\dd{p}}{2g}(u^2+v^2)\dd{A}.
\end{equation}
The kinetic energy of the model runs takes on the order of $>15000$
days to reach a steady-state (note unless
otherwise specified, a ``day'' is an \textit{Earth} day, i.e. 86400
seconds). With a dynamical timestep of ~100 s required for stability,
the wall-time duration for 1000 days of simulation was between 1-2 days
(running the model across 24 CPUs). On the
other hand, the maximum and minimum zonal wind of the model reach a
steady state on day $\approx 10000$ for the 33 day period runs, and
$< 5000$ days for the 6 day period runs. Whilst the kinetic energy
is dominated by the denser lower atmosphere, the jet speeds are
maximal in the upper atmosphere which takes a shorter period of time
to reach its steady state. In the 33 day period experiments,
  we see the formation of equatorial superrotation in the upper
  atmosphere only after 10000 days (see
  section~\ref{subsec:superrotation}). The globally-integrated outgoing longwave
radiation (OLR) of the model reaches equilibrium with the incoming
stellar radiation after $\approx 1000$ days. We see temporal
  variability of the total kinetic energy even in the equilibrium
  state. This is caused by fluctuations in the wind field due to the
  non-linearity of the equations being solved, and also due to the
  recurrence of an instability in the upper atmosphere (see
  section~\ref{subsec:superrotation}) which results in variation of
  the wind field with time. We note that the P6d experiment may not be
in equilibrium, with large variations in kinetic energy even at 20000 days. In \cite{Wang2020}
the convergence of the lower atmosphere took on the order of $10^5$
days, after which the kinetic energy of the atmosphere reached a
steady-state value. Our timescale appears to be shorter, which could
be attributed to two things: firstly, our model has a lower bottom
boundary pressure of \SI{10}{bar} so the radiative timescale will be
shorter at the bottom of our model; secondly, we initialised the
simulations with the vertical temperature profile equal to the
analytic radiative equilibrium temperature of a 1D grey gas model. As
shown in section~\ref{subsec:tempstructure}, since there are extremely weak
horizontal temperature gradients throughout the atmosphere, 3D
vertical temperature profiles fit the analytic profile well. This
would reduce the length of time the lower atmosphere has to adjust to
radiative equilibrium (compared to initialising on an isothermal
profile, as done in \cite{Wang2020} and previous studies of GJ~1214b). 
  \subsection{Temperature Structure} \label{subsec:tempstructure}
  
\begin{figure*}
  \epsscale{1.1}
  \plotone{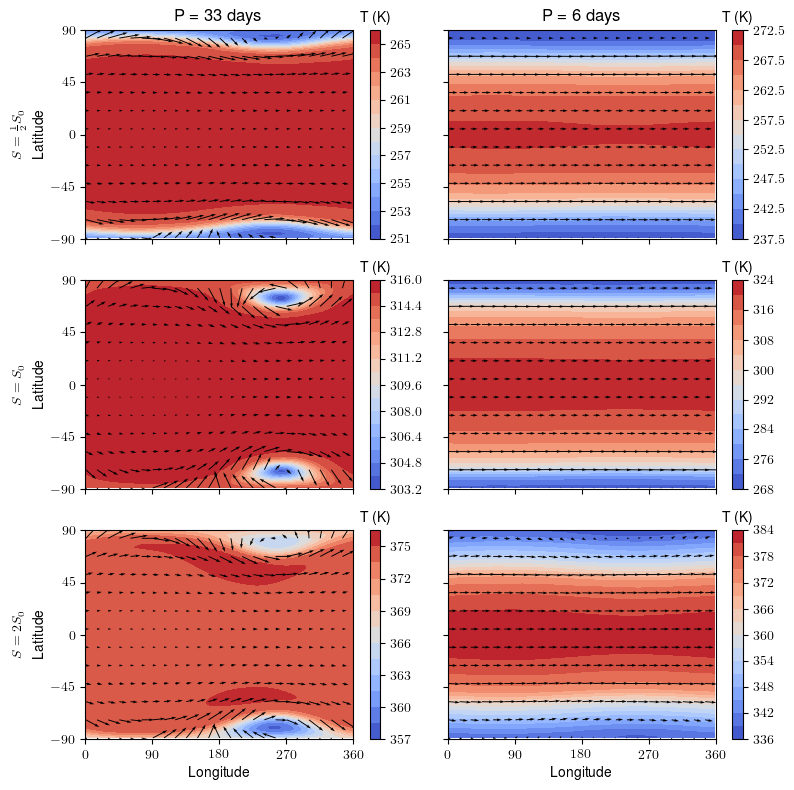}
  \caption{The \SI{100}{mbar} level temperature profiles for the dry
    experiments. For the 33 day period experiments, the temperature
    varies on the order of 1 K throughout most of the horizontal plane
    apart from at a cold spot near (but offset from) the poles. For the 6 day period experiments,
    there is a much larger temperature drop from equator to
    pole. Note that the substellar point is at $0^{\circ}$ longitude. \label{fig:temp}}
\end{figure*}

\begin{figure*}
  \epsscale{1.1}
  \plotone{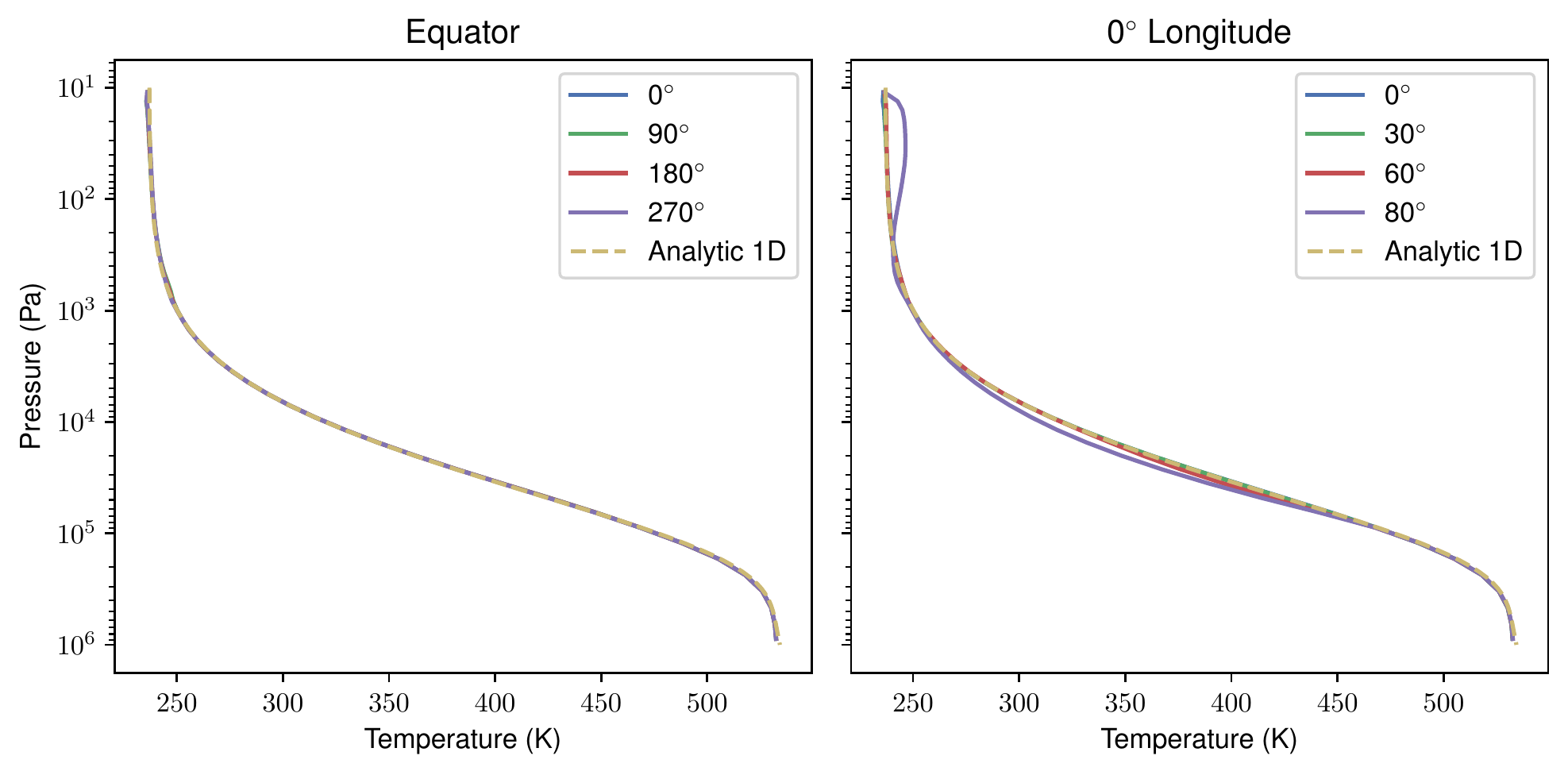}
  \caption{Vertical temperature profiles for the control
    experiment. Left: On equator, the temperature profiles are very
    uniform and conforming to the analytic 1D solution. Right: In the
    latitudinal direction, there are deviations from uniform
    temperature gradients, but only near the poles. \label{fig:temp2}}
\end{figure*}

Figure~\ref{fig:temp} shows the horizontal temperature structure at \SI{100}{mbar}. In all plots the substellar point is
defined to be at $(\phi,\lambda) = (0^{\circ}, 0^{\circ})$. The slower rotating experiments (with $P=$ 33
days) show extremely horizontally uniform temperature profiles, with
the temperature varying on the order of 1 K apart from near the poles,
where there is a more significant drop in temperature in cold,
stationary vortex-like structures. In contrast,
the faster rotating experiments show a much larger temperature drop
between the equator and pole. The faster-rotating experiments support stronger cyclostrophic jets than the 33 day period experiment (see sections~\ref{subsec:zonal_circ} and~\ref{subsec:overturning}) which in turn require a meridional temperature gradient to sustain by gradient wind balance. The slower-rotating experiments are also
less uniform zonally, especially near the poles where there is a
region where the meridional and zonal winds are of the same
magnitude. In the 6-day period experiments, however, the wind field is
extremely zonally uniform at all latitudes.
The vertical temperature profiles in all runs closely fit the analytic
1D temperature profile assuming perfect heat redistribution, apart
from near the poles. Figure~\ref{fig:temp2} shows vertical temperature
profiles at different latitudes and longitudes and in comparison to
the 1D analytic solution for the control
experiment. Appendix~\ref{app:b} shows the same data for the
  rest of the experiments. On the equator, in the zonal direction there
is very little temperature variation. Moving polewards from the
substellar point, however, there are significant variations in
temperature away from radiative equilibrium.  
The weak temperature gradients in the tropics and mid-latitudes can be
explained from a simple scaling argument, following
\cite{Pierrehumbert2016}. The momentum equation, written in
non-dimensionalised form is:
\begin{equation}
  \label{eq:wtg}
  \text{Ro}^2(\boldsymbol{u}\cdot\grad u - uv\tan\phi) -
  \text{Ro}\cdot fv = -\Lambda^2\partial_x\Phi,
\end{equation}
where Ro is the Rossby number, $U/(\Lambda a)$, and $\Lambda$ is the
weak temperature gradient (WTG) parameter $c_0/(\Lambda a)$
\citep{Pierrehumbert2016,Pierrehumbert2019} which is the ratio of the
Rossby deformation radius to the planetary radius. $c_0$ is a characteristic wave speed. If we approximate the
atmosphere as isothermal with characteristic temperature $T$, this
wave speed $\sim \sqrt{RT}$. For weak horizontal temperature
gradients, we require $\Lambda\gg 1$ and $\text{Ro}\gg 1$ such that
$\partial_x\Phi \sim \partial_x T \approx 0$. Using estimates from our
experiments, the 33 day period simulations have $\text{Ro}\approx 10$,
$\Lambda\approx 30$, whilst the 6 day period simulations have
$\text{Ro}\approx 3$, $\Lambda\approx 10$, confirming that both cases
are firmly within the WTG regime, with the slower-rotating experiments
expected to be more strongly uniform in temperature. In contrast,
temperate terrestrial simulations (such as \cite{Pierrehumbert2019})
often have high $\mu$ atmospheres, making $\Lambda < 1$ and exciting a
strong wave response. GJ 1214b, being much hotter and faster rotating,
has $\Lambda\approx 1$, being greater than or less than 1 depending on the assumed metallicity of its
atmosphere (see section~\ref{subsec:comp}).

This result can also be interpreted using the theory from
\cite{Zhang2017} in the low-drag limit, where the day-night temperature contrast compared to
the equilibrium contrast is given by:
\begin{subequations}
\begin{align}
  \label{eq:3}
  \frac{\Delta T}{\Delta T_{\text{eq}}}&\sim 1 - \frac{2}{\alpha +
                                         \sqrt{\alpha^2 + 4\gamma^2}} \\
  \alpha &= 1 + \frac{\Omega\tau_{\text{w}}^2}{\tau_{\text{r}}\Delta\ln p}\\
  \gamma &= \frac{\tau_{\text{w}}^2}{\tau_{\text{r}}\tau_{\text{a,eq}}\Delta \ln p}
\end{align}
\end{subequations}
where $\tau_{\text{r}}$, $\tau_{\text{w}}$ and $\tau_{\text{a,eq}}$
are the radiative, wave and cyclostrophic advective timescales
respectively \citep{Zhang2017}. If we write the wave timescale as
$a/\sqrt{RT} = 1/(\Omega\Lambda)$ and the advective timescale as:
\begin{align}
  \label{eq:adv}
  \tau_{\text{a,eq}} &= \frac{a}{U_{\text{eq}}}\\
                     &= \frac{a}{\sqrt{R\Delta T_{\text{eq}}\ln p}}\\
  & = \frac{1}{\kappa \Lambda\Omega\sqrt{\ln p}},
\end{align}
where $\kappa^2\ll 1$ is the ratio between the equilibrium day-night
temperature difference and the characteristic temperature of the
atmosphere, then:
\begin{align}
  \label{eq:alphagamma}
  \alpha &= 1 + \frac{1}{\tau_{\text{r}}\Omega\Lambda^2\Delta\ln p}\\
  \gamma &= \frac{\kappa\sqrt{\ln p}}{\Lambda\tau_{\text{r}}\Omega}.  
\end{align}
The radiative timescale at the $\tau=1$ level can be estimated as:
\begin{equation}
  \label{eq:taur}
  \tau_{\text{r}} = \frac{c_pp(\tau=1)}{4\sigma T^3g} =
  \frac{c_pp_s}{4\sigma T^3g\tau_{\text{LW}}}\approx 8~\text{days}
\end{equation}
which means $(\tau_{\text{r}}\Omega)^{-1} = 1$ for the $P=33$ day
experiment and 0.1 for the $P=6$ day experiments. Therefore in all our
experiments we predict $\alpha\approx 1$ and $\gamma\ll 1$, which
corresponds to $\Delta T/\Delta T_{\text{eq}}\ll 1$.
\subsection{Zonal Circulation} \label{subsec:zonal_circ}

\begin{figure*}
  \plotone{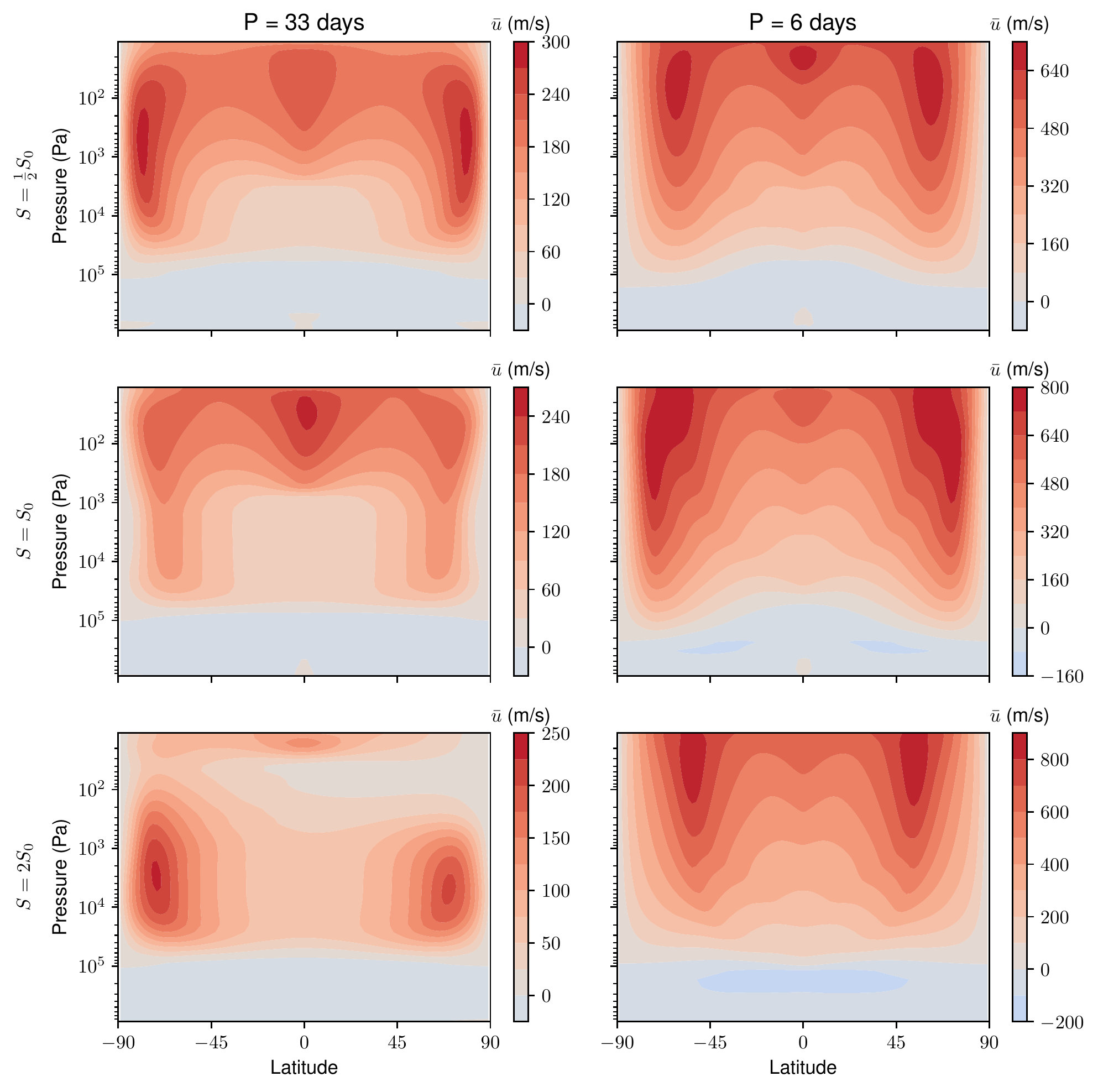}
  \caption{The zonal mean zonal wind for each of the experiments. All circulations exhibit high latitude cyclostrophic
    jets. All experiments show equatorial super rotation, in some
    cases forming a third maximum on-equator. \label{fig:zwind} }
\end{figure*}

\begin{figure*}
  \epsscale{0.92}
  \plotone{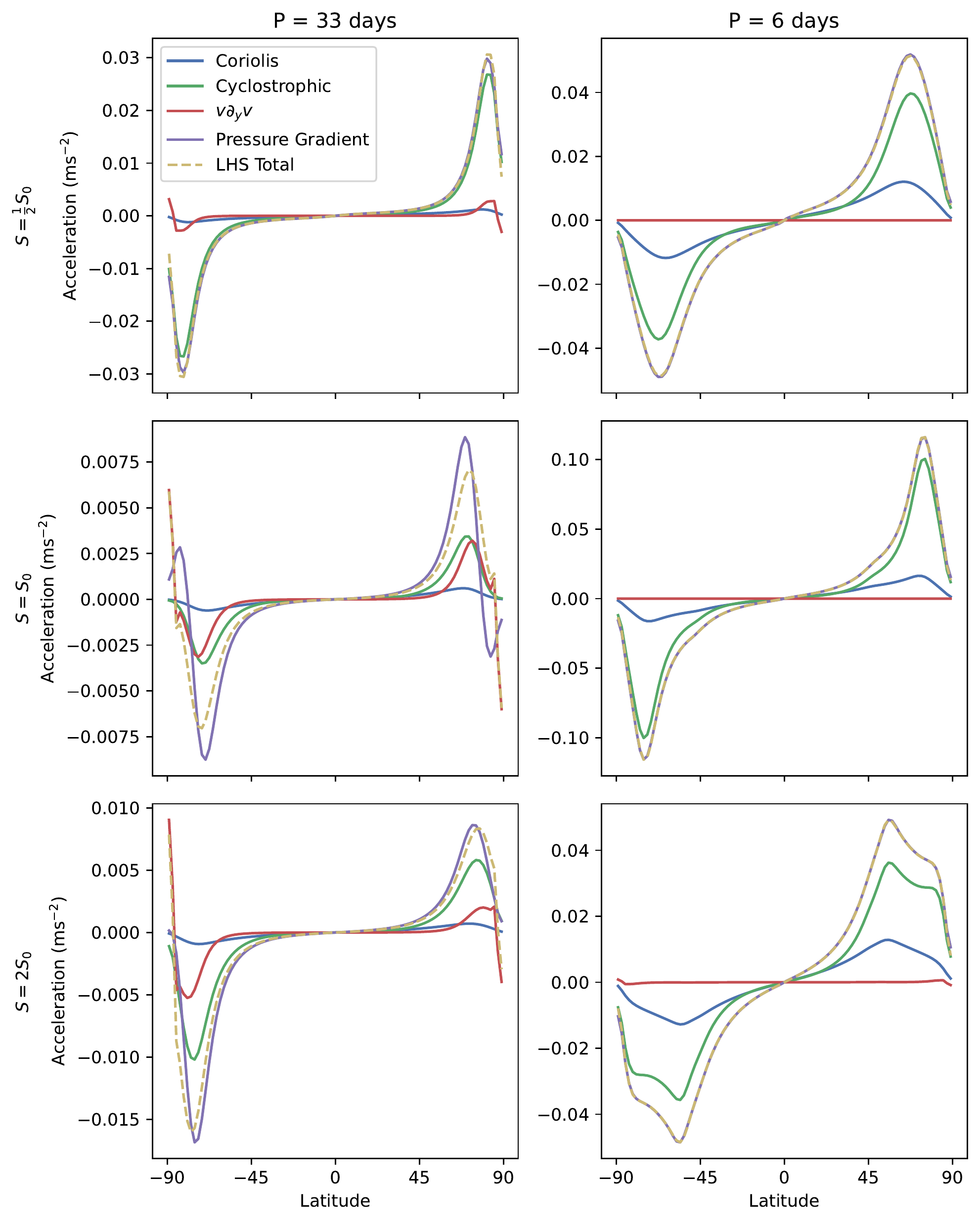}
  \caption{Meridional momentum balance for each of the dry
    simulations at the 10 mbar level. For the six-day period
    experiments (right-hand side), classical gradient-wind balance
    sustains the jets, whereas in the 33 day period experiments the
    non-linear advection term is also significant to the balance. \label{fig:bal}}
\end{figure*}

Next, we look at the zonal circulation. The
temperature and wind profiles in our model runs were largely zonally
symmetric, meaning the zonally averaged circulation is a good way of
reducing the dimensions of the output data. Figure~\ref{fig:zwind}
shows the zonal mean zonal wind of each experiment. 
The common feature of all the circulations is the presence of two
high-latitude cyclostrophic jets which are in steady-state
balance. Figure~\ref{fig:bal} shows the non-negligible terms in the
zonally averaged meridional momentum equation:
\begin{equation}
  \label{eq:balance}
  \frac{1}{a}\qty[v\partial_{\phi}v] + \frac{\qty[u^2]\tan\phi}{a} +
  \qty[fu] = -\frac{1}{a}[\partial_{\phi}\Phi],
\end{equation}
where $[\cdot]$ represents zonally averaged terms. The experiments
with the shorter period orbit are very close to gradient wind balance,
with the pressure gradient induced by the temperature contrast between
the equatorial regions and the polar regions balanced by the Coriolis
forces and cyclostrophic terms in the momentum equation. In contrast,
in the slower-rotating experiments, non-linear advective terms become
important near the poles, where the meridional wind reaches its
maximum magnitude value. Table~\ref{tab:merid} shows the maximum value
of the time-averaged meridional velocity in each of the experiments,
from which it is clear that the meridional velocity is an order of
magnitude stronger in the slower-rotating experiment.
\begin{deluxetable}{ll}
  \tablehead{\colhead{Experiment name} & \colhead{Maximum $v$ (m/s)}}
  \tablecaption{Maximum meridional velocity in the time-averaged flow \label{tab:merid}}
  \startdata
  PKh & 193 \\
  PKc & 265 \\
  PKd & 383 \\
  P6h & 7 \\
  P6c & 13 \\
  P6d & 29
  \enddata
\end{deluxetable}

We do not currently have a theory for predicting the strength of this
zonal circulation. The gradient wind balance shown in~\ref{fig:bal} is
a diagnostic balance which does not illuminate how the atmosphere
arrives in its final state. In other GCM studies of tidally-locked
exoplanets
\citep{Komacek2016,Zhang2017}, the RMS wind is linked to the
advective, wave and radiative timescales. However, in these studies
the wind speed also depends on an equilibrium temperature profile
which is prescribed in the model. In our grey gas model, this
equilibrium temperature profile is not known a priori which makes it
difficult to produce similar scaling relations. We note that at each rotation rate, the
zonal jet strength is approximately the same at each instellation, but
that it increases greatly with increasing rotation rate. An angular
momentum conserving wind moving from the equator with $u=0$ would
have speed:
\begin{equation}
  \label{eq:17}
  u_{\text{AM}} = a\Omega\frac{\sin^2\phi}{\cos\phi}
\end{equation}
at latitude $\phi$. For jets at $\phi=75^{\circ}$, this gives
$u_{\text{AM}} = 130$ m/s for the 33 day period case and
$u_{\text{AM}} = 730$ m/s for the 6 day period case, which is of the
correct order of magnitude as the results in Figure~\ref{fig:zwind}. However,
we note that there is no requirement for the zonal wind speed to be 0
at the equator (as in a classical Hadley circulation) since there is
no friction at the bottom of the model. We investigate the overturning
circulation in the next section. 

\subsection{Overturning Circulation}
\label{subsec:overturning}

\begin{figure*}
  \plotone{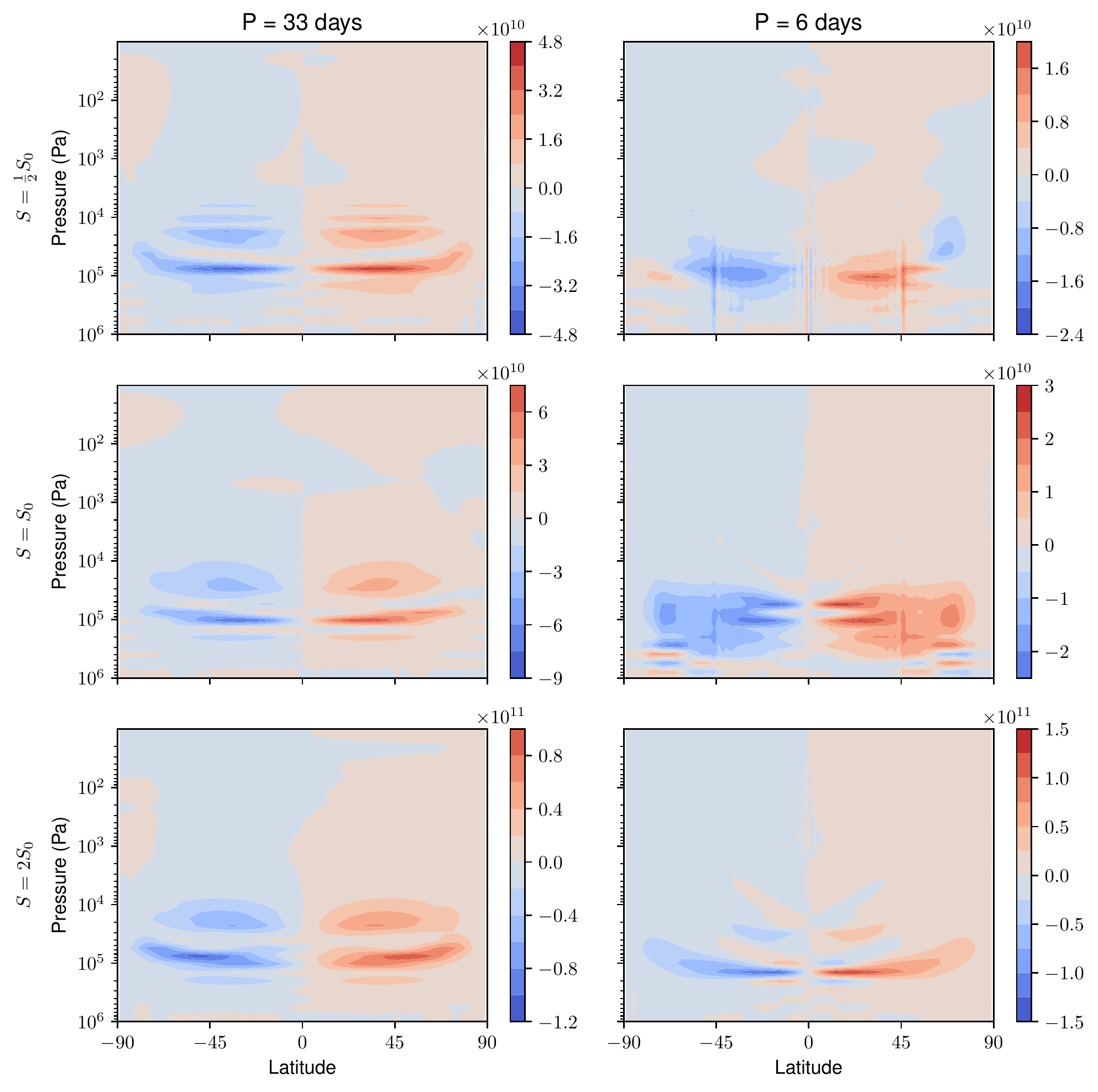}
  \caption{The meridional mass streamfunction (contours in units of kg/s). All the experiments show
    some form of net overturning from the equator to pole. \label{fig:mpsi}}
\end{figure*}

To investigate the origin of the high-latitude zonal jets, we look at
the mean meridional overturning circulation. In the conventional
$(\lambda, \phi)$ coordinate system, the mass streamfunction is given
by:
\begin{equation}
  \label{eq:psi}
  \psi_m(\phi,p) = \frac{2\pi a\cos\phi}{g}\int_0^p[v](\phi,p')\dd{p'}, 
\end{equation}
which represents the meridional mass flux between the top of the
atmosphere ($p=0$) and a pressure level $p$. Figure~\ref{fig:mpsi}
shows this mass streamfunction all the experiments. The
zonal-mean circulation follows contours of constant $\psi_m$, so
positive (red) regions indicate clockwise circulation around contours,
and negative (blue) regions anticlockwise circulation. In all six cases we see
a net overturning circulation from equator to pole in both
hemispheres. In the 33 day period experiments, we also see two
separate cells appear in each hemisphere, with one centered at
approximately \SI{2e4}{Pa}
and another centered at around \SI{1e5}{Pa}. In the theory of non-synchronously rotating planets,
having $\text{Ro}\gg 1$ leads to a global Hadley-like circulation
(e.g. Venus and Titan), which
transports energy from the tropics to the poles to balance the
gradient in stellar heating \citep{Held1980,Vallis2017,Kaspi2015}. This Hadley circulation also transports
angular from the equator polewards and spins up sub-tropical jets on
Earth. Our experiments differ from the classical Held-Hou model in two
main ways: firstly, we do not have a drag layer at the bottom of the
atmosphere which can inject angular momentum into the atmospheric flow
by imposing a torque on westwards moving surface-flow. Our experiments
conserve the integrated angular momentum of the atmosphere to one part
in 1000 over the total simulation time (with losses due to numerical
dissipation). Secondly, the forcing in our model is non-axisymmetric
and is centred on the substellar point, meaning the main heating
contrast is between substellar and antistellar points in contrast to
between equator and pole, as in the canonical analysis.

Since there is no surface to inject angular momentum into our atmosphere, the positive angular momentum associated with jets in
the upper atmosphere is balanced by a retrograde circulation in the
lower atmosphere. Spinning up from rest, the radiative heating at the
equator causes rising motion there. Air moving polewards higher up in
the atmosphere rotates in the prograde direction, conserving angular
momentum, whereas the return flow will have retrograde motion in the
absence of friction. How these two circulations interact is unclear
and requires more study. As mentioned above,
since $u$ is not fixed at 0 at the equator, it is difficult to use
this as a theory to predict the jet speed away from the
equator. However, the qualitative picture of air from the equator
moving polewards and accelerating the jets provides a rough
order-of-magnitude estimate of the jet wind speeds.

\begin{figure*}
  \plotone{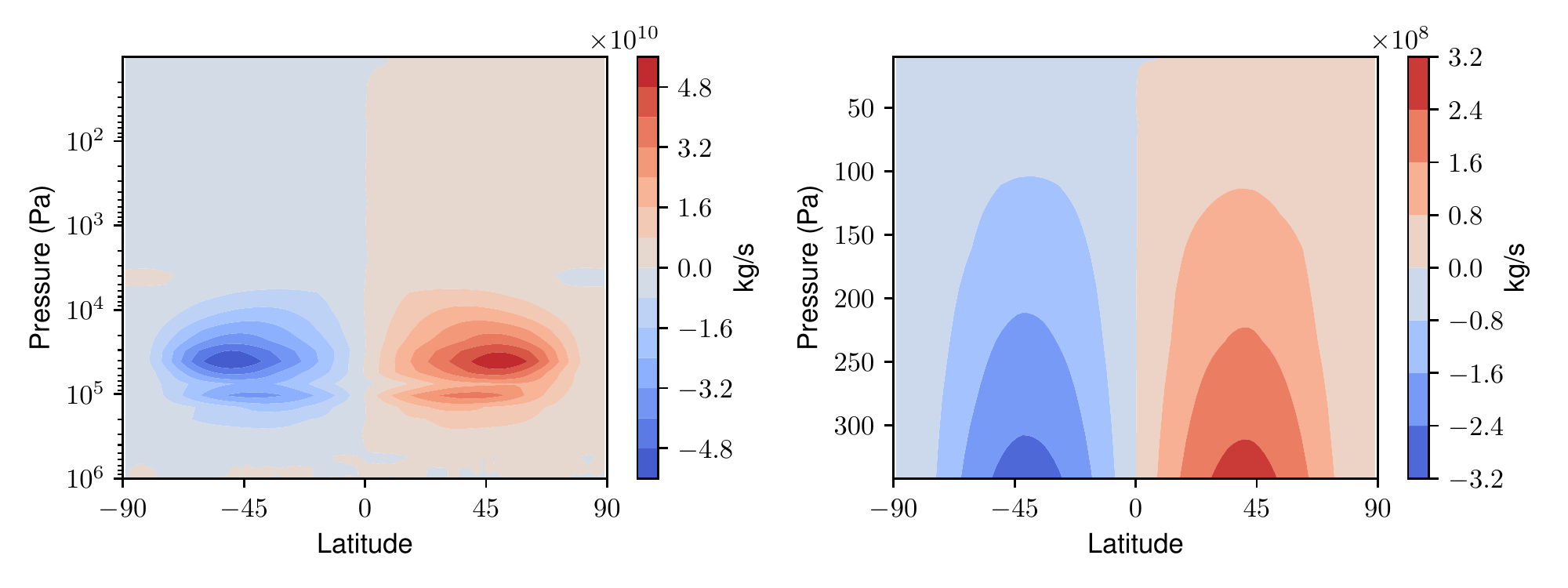}
  \caption{The mass streamfunction of the control experiment averaged
    between 4000 and 5000 days during spin-up. Left: A view of the
    the whole atmosphere with it's clear overturning circulation from
    equator to pole. Right: The upper atmosphere's streamfunction has
    similar structure (though is obscured on the left due to the
    magnitude), highlighting that the upper atmosphere is spun up via
    a similar mechanism. \label{fig:sf_toa}}
\end{figure*}

Figure~\ref{fig:sf_toa} shows the mass streamfunction of the control
experiment during spin-up (averaged between 4000 and 5000 days). The
overturning circulation that would spin up angular momentum conserving
jets is clear on the left-hand panel. The picture in the upper
atmosphere is similar (but the overturning circulation is weaker due
to the lower mass of the upper atmosphere)

\begin{figure*}
  \plotone{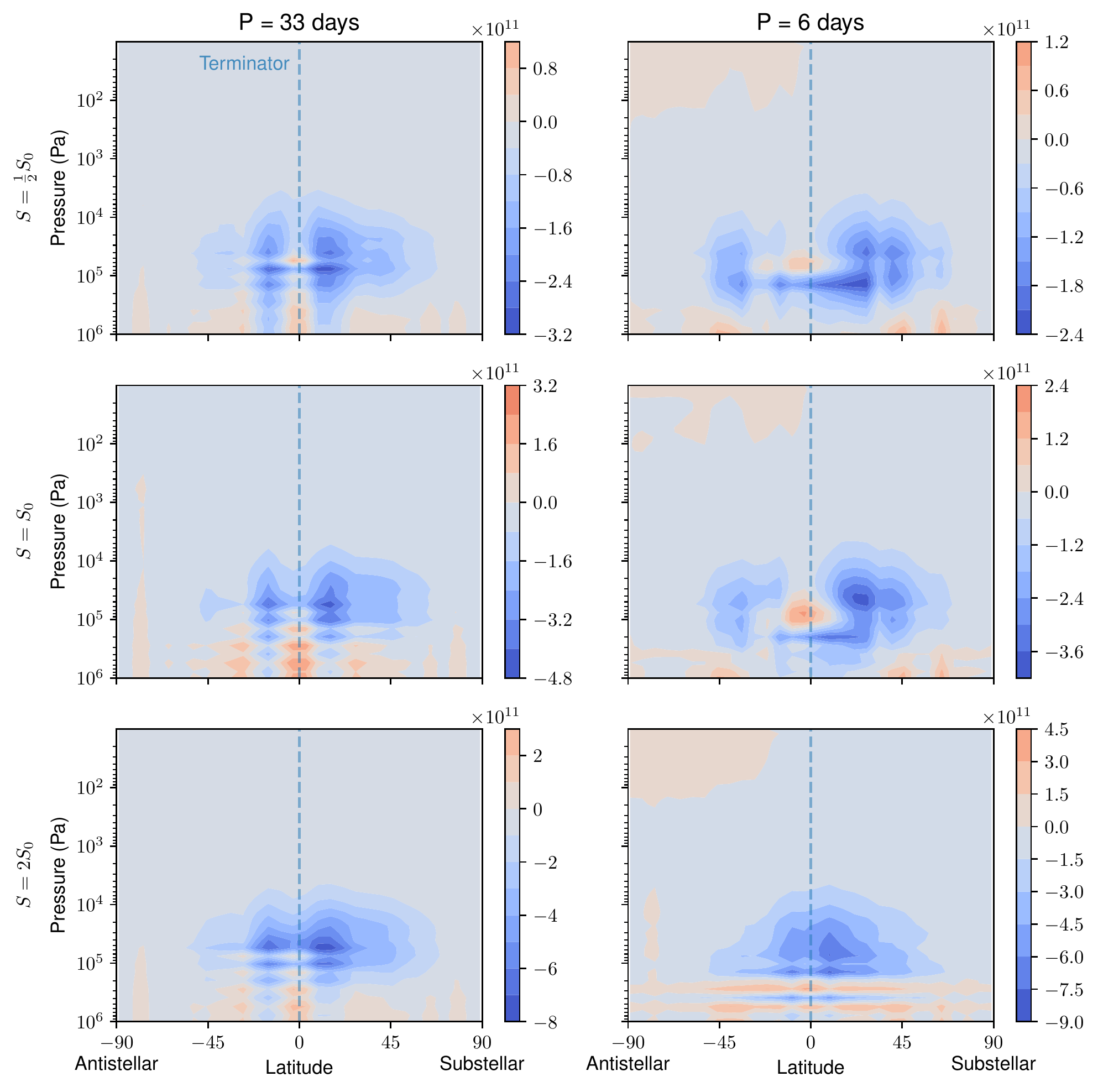}
  \caption{The tidally-locked mass streamfunction, $\psi_{\text{TL}}$
    (units kg/s). This shows more clearly the stronger
    overturning circulation between dayside and nightside (c.f. Figure~\ref{fig:mpsi})\label{fig:tlpsi}}
\end{figure*}

In previous
studies of tidally-locked planets, the zonal circulation has sometimes
been split into a dayside Hadley circulation and a nightside
anti-Hadley circulation \citep[e.g.][]{Charnay2015a} -- however, this
can obscure a more direct circulation between dayside and nightside
\citep{Hammond2021}. We therefore follow \cite{Hammond2021} in
calculating the mass streamfunction in tidally-locked (TL) coordinates
\citep{Koll2015}. The TL latitude, $\phi_{\text{TL}}$,
is 0 degrees at the North and South poles, 90 degrees at the
substellar point and -90 degrees at the antistellar point. The TL longitude, $\lambda_{\text{TL}}$, is 0 on the semi-circle passing
through the substellar point, North Pole and antistellar point and
increases towards the Eastern terminator. To avoid ambiguity, we will
use the term ``the poles'' to refer exclusively to the poles in the
conventional latitude-longitude coordinate system, and use
``substellar point'' and ``antistellar point'' to refer to the points
at $\phi_{\text{TL}}=90^{\circ}$ and $\phi_{\text{TL}}=-90^{\circ}$
respectively. In TL coordinates,
one can define a streamfunction:
\begin{equation}
  \label{eq:5}
  \psi_{\text{TL}} = \frac{2\pi a \cos\phi_{\text{TL}}}{g}\int_{0}^{p}[v_{\text{TL}}]_{\text{TL}}(\phi_{\text{TL}},p')\dd{p}
\end{equation}
where $[\cdot]_{\text{TL}}$ represents an average over
$\lambda_{\text{TL}}$, and $v_{\text{TL}}$ is the component of the
wind in the direction of increasing $\phi_{\text{TL}}$ (and not just
the conventional meridional wind in a different coordinate
system). Figure~\ref{fig:tlpsi} shows the streamfunction in this
coordinate system. In all experiments there is a direct day-night
circulation driven by radiative heating. In the six-day period
experiments, there is also a small area of counter-circulation near
the terminator. This is because the jet regions near the poles are associated with a
downwelling region on the the dayside and upwelling on the nightside. 
One can also see, in all six cases, that the flow is partially splits
into two sub-circulations, with a stronger one on the dayside and a
weaker one on the nightside. Thinking in terms of the conventional
mass streamfunction, this would explain why we see a circulation from equator to pole -- because the ``anti-Hadley''
circulation on the nightside is weaker than the conventional
Hadley-like circulation on the dayside (due to weaker radiative
forcing on the nightside). This asymmetry between the dayside and nightside flow allows for a net transport of angular momentum from equator to pole, which accelerates the zonal jets.

We note that this circulation is on average an order of magnitude
larger than the overturning circulation in latitude-longitude
coordinates, confirming that this is the dominant overturning
circulation. 

To better distinguish the overturning circulation from the zonal jet
structure, we decompose the horizontal winds into their divergent and
rotational components \citep{Hammond2021}:
\begin{subequations}
\begin{align}
  &\boldsymbol{u} = \boldsymbol{u}_{\text{div}} + \boldsymbol{u}_{\text{rot}}\\
  &\boldsymbol{u}_{\text{rot}} = \boldsymbol{k}\times \grad\Psi \\
  &\boldsymbol{u}_{\text{div}} = -\grad{\chi}
\end{align}
\end{subequations}
where $\Psi$ is a streamfunction (distinct from the mass
streamfunctions discussed) and $\chi$ is a velocity potential
which can be calculated from the vorticity and divergence of the
velocity respectively. Importantly, the
meridional rotational velocity is given by $v_{\text{rot}} =
\partial_x\Psi$, which vanishes when averaged in the longitudinal
direction. Therefore, the mass streamfunction depends only on the
divergent component of the velocity field. This is true in both
conventional latitude-longitude and TL coordinate systems.
\begin{figure}
  \plotone{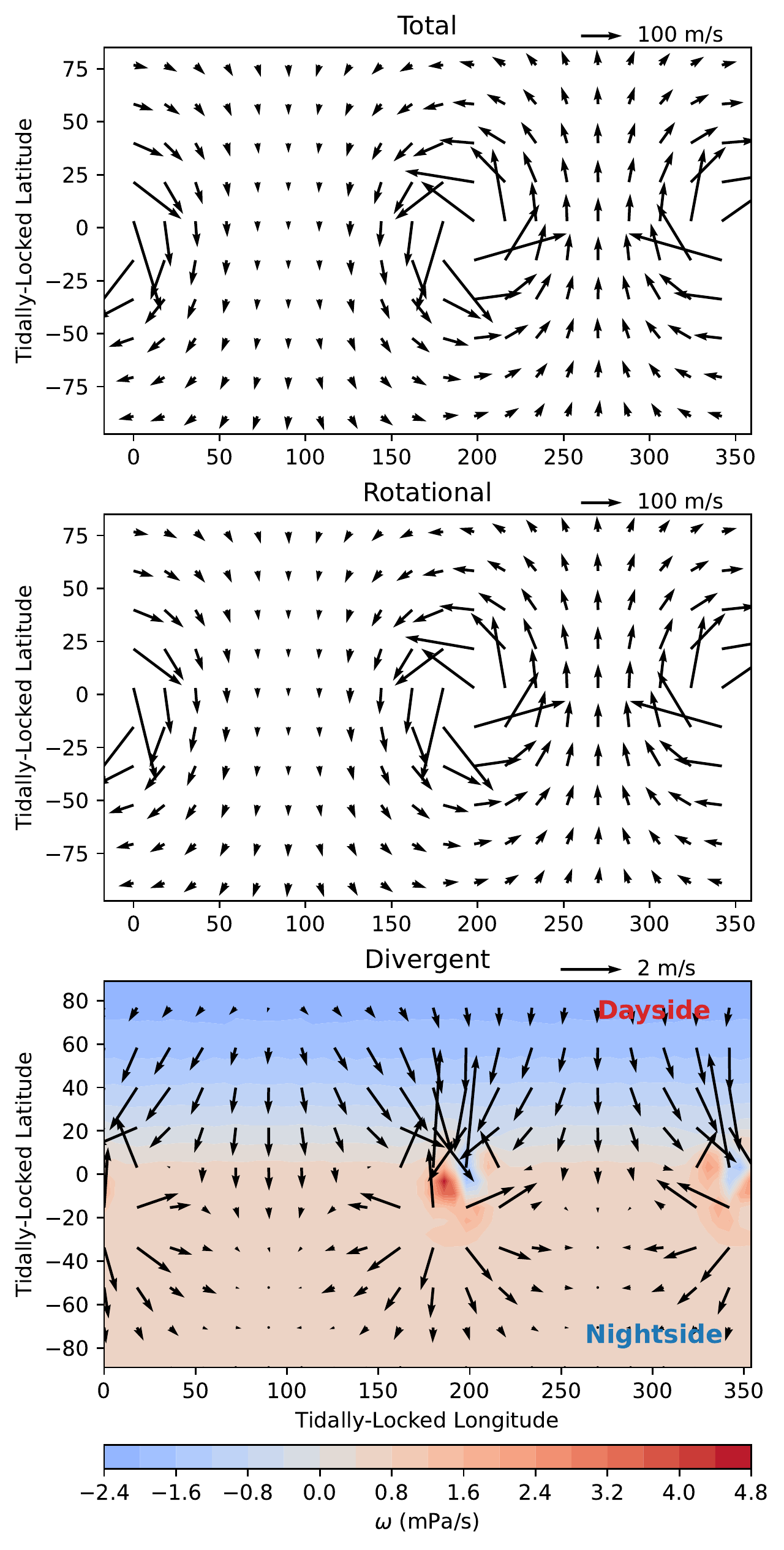}
  \caption{Total, rotational and divergent components of the wind
    field of the control experiment at 100 mbar. The rotational
    component of the wind dominates the circulation, but the small
    divergent component is responsible for the mass transport between
    dayside and night-side. The bottom plot also shows the pressure
    velocity, $\omega$, which illustrates the upwelling on the
    dayside and downwelling on the nightside of the planet. \label{fig:divrot}}
\end{figure}
Figure~\ref{fig:divrot} shows the total, rotational and divergent
winds for the control experiment at 100 mbar, along with the pressure
velocity. We find that the velocity is dominated by the rotational
component with magnitude on the order of 100 m/s. However, the
divergent component responsible for the overturning circulation, shown
in the bottom of Figure~\ref{fig:divrot} is predominantly directed
from dayside to nightside (with some erratic behaviour at the
conventional North and South poles, most likely due to regridding errors).

We estimate the dependence of the magnitude of the TL streamfunction on
instellation using the steady-state equation for the transport of dry static
energy, $s\equiv c_pT + \Phi$:
\begin{equation}
  \label{eq:dse1}
  \div (s\boldsymbol{u}_{\text{TL}}) = \dot{Q},
\end{equation}
where $\dot{Q}$ is the heating rate per unit mass. Taking the average over the TL longitude and integrating
over $\dd{p}/g$, we find:
\begin{equation}
  \label{eq:tlsf}
  \int_0^{p}\frac{\dd{p}}{g}\frac{1}{a\cos\phi}\partial_\phi [v_{\text{TL}}s\cos\phi] = [S-F],
\end{equation}
where $S$ is the pressure-integrated solar heating and $F$ is the
pressure-integrated outgoing flux. We have also used the fact that $\omega s = 0$
at both $p=0$ (true from the boundary condition $\omega=0$ there), and
at the level and latitude of the maximum of the TL streamfunction (also
true since $\omega\sim\pdv*{\psi}{\phi}=0$ where $\psi_{\text{TL}}$ is
maximal). We now
approximate this equation to find a scaling balance of terms by making
a series of approximations. Firstly, we assume that the dry static
energy and $v_{\text{TL}}$ are uncorrelated in the zonal and vertical
directions. We also assume that if the temperature gradients are weak
enough, then $\Delta s/s \ll \Delta v/v$ such that:
\begin{align}
  \label{eq:dse2}
  \int_0^{p}
  \frac{\dd{p}}{g}\frac{1}{a\cos\phi}\partial_\phi[v_{\text{TL}}s\cos_\phi]\\
  \approx \langle[s]\rangle\int_0^{p}\frac{\dd{p}}{g}\frac{1}{a\cos\phi}\partial_{\phi}({[v_{\text{TL}}]\cos\phi})\label{eq:dse3}
\end{align}
where $\langle \cdot \rangle$ represents a column average. Note that
the integral term in~\ref{eq:dse3} is proportional to $\omega$ which
is in term proportional to $\partial_{\phi}\psi_{\text{TL}}$. To
approximate $[S-F]$, we note that when $p\to p_s$, $S$ becomes the
incoming stellar radiation and $F$ the OLR, which vary over the scale
$S_0$ in the $\phi_{\text{TL}}$ direction. We approximate that this scaling holds at the level of the maximum
streamfuncton too. Thus
we get the scaling:
\begin{equation}
  \label{eq:psiscale}
  \psi_{\text{TL}} \sim \frac{a^2S_0}{\langle [s] \rangle}
\end{equation}
The dry static energy, $s=c_pT + \Phi$, scales as $\max(c_pT,
\Phi)\sim T\sim S_0^{1/4}$ if we assume $\Phi\approx RT$ i.e. the
region integrated over is roughly one scale height. This gives the
final scaling $\psi_{\text{TL}}\sim S_0^{3/4}$. Note that there is no
dependence on the rotation rate in this expression. 

\begin{figure*}
  \gridline{\fig{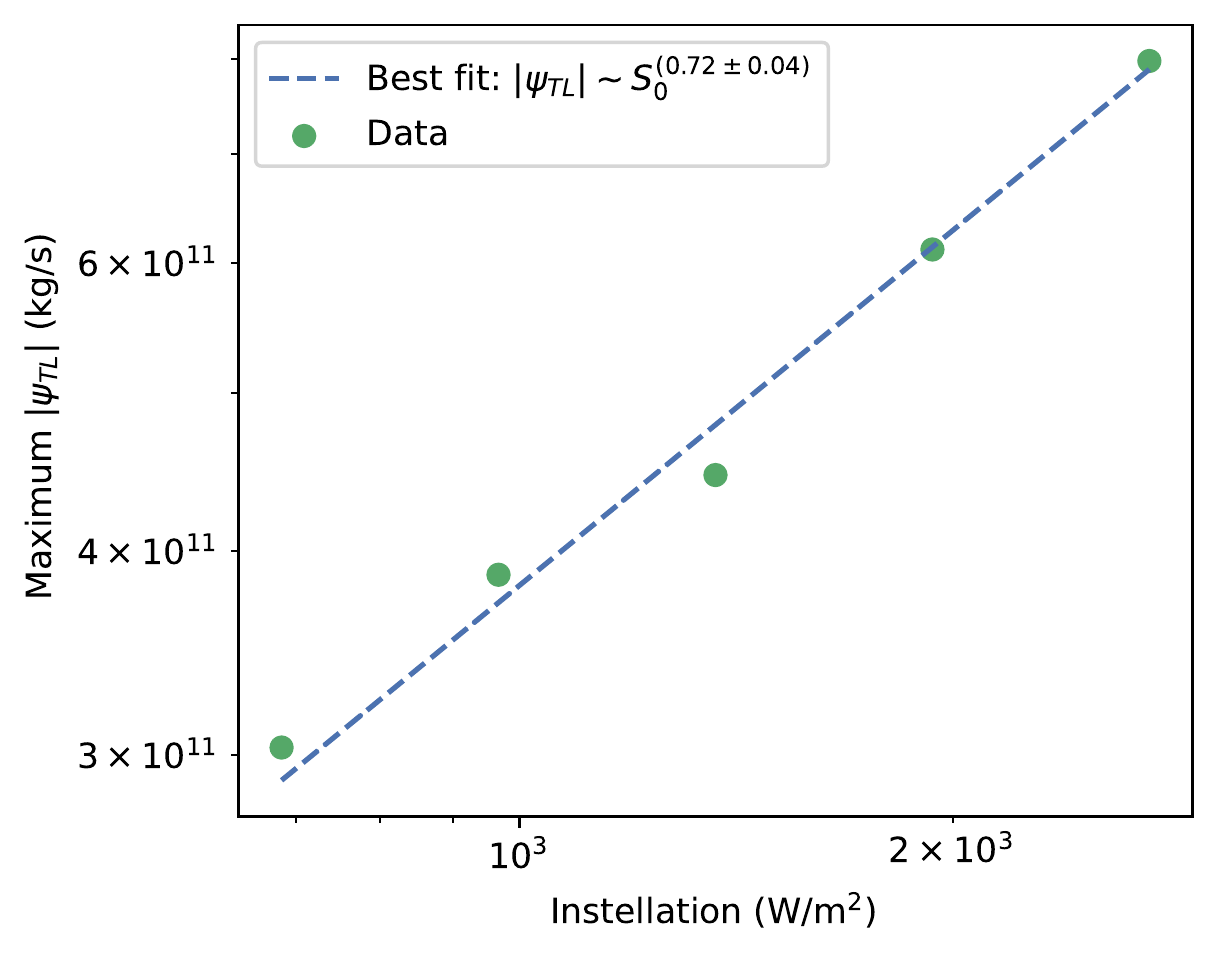}{0.5\textwidth}{(a)}
    \fig{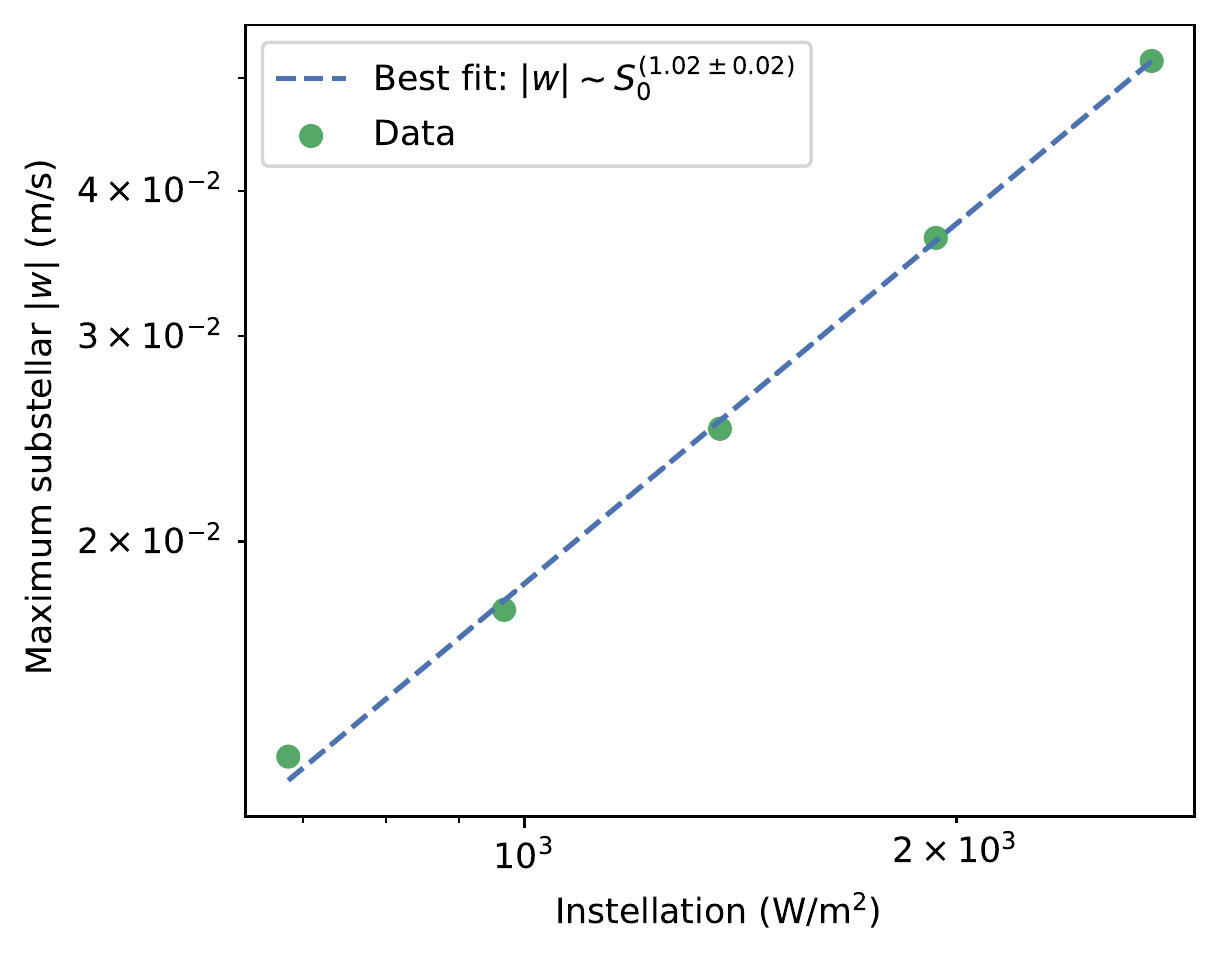}{0.5\textwidth}{(b)}}
  \caption{(a) The maximum absolute value of the tidally-locked
    streamfunction as a function of instellation. The best fit power
    law to the data shows $\psi_{\text{TL}}\sim S_0^{0.72\pm0.04}$,
    which is in agreement with the theoretical exponent,
    0.75. (b) The maximum value of the vertical velocity, $w$, at the
    substellar point as a function of instellation. The best fit
    power law to the data shows $w\sim S_0^{1.02\pm 0.02}$ which is
    in agreement with the theoretical exponent, 1. \label{fig:sf_inst} }
\end{figure*}

Figure~\ref{fig:sf_inst}(a) shows the maximum absolute value of the
TL streamfunction from five experiments at
different instellations. The best-fit power law relation has exponent
$0.72\pm 0.04$, which is in agreement with the theoretical prediction
of $\frac{3}{4}$.

We can also use this theory to predict the characteristic vertical
velocity of the overturning circulation. The vertical velocity, $w =
\dv*{z}{t}$, is related to the pressure velocity $\omega$ by:
\begin{equation}
  \label{eq:9}
  w = -\frac{1}{\rho g}\omega = -\frac{RT}{pg}\omega.
\end{equation}
The pressure velocity is linked to the streamfunction by:
\begin{equation}
  \label{eq:10}
  \omega =  -\frac{g}{2\pi a^2\cos\phi}\pdv{\psi_{\text{TL}}}{\phi}.
\end{equation}
Combining these two equations and using $\partial_\phi\psi_{\text{TL}}\sim\psi_{\text{TL}}\sim
a^2S/\langle [s] \rangle$ we get:
\begin{equation}
  \label{eq:11}
  w \sim \frac{S_0}{p}.
\end{equation}
Note this scaling assumes that the pressure level of maximum vertical
velocity is the same as the pressure level of the maximum
streamfunction. In all our experiments, although $S_0$ is varied, the
grey opacities stay the same, so the level of maximum radiative
heating (linked to the level of maximum upwelling by the WTG
approximation) stays approximately the same. We find that the maximum
value of $w$ in all experiments is at $p\approx$ \SI{2e4}{Pa}, in which
case $w\sim S_0$. Figure~\ref{fig:sf_inst} confirms this dependence,
with the best-fit line having an exponent of $1.02\pm 0.02$, in
agreement with theory.

\subsection{Equatorial Superrotation}\label{subsec:superrotation}
In each of the experiments shown in Figure~\ref{fig:zwind}, there is
equatorial superrotation. There are two main types of mechanism
thought to drive superrotation in planetary atmospheres
\citep{Wang2014}. The first is through generating eddies from
non-axisymmetric forcing, e.g. via tidally-locked solar heating
\citep{Showman2011,Tsai2014,Hammond2018} or from convective heating
\citep[e.g.][]{Lian2010,Liu2011}. The second is from shear
instabilities in atmospheres with axisymmetric forcing \citep[e.g.][]{Iga2005}, which is
observed in idealized simulations of slow-rotating planets \citep{Mitchell2010} and proposed as a potential
mechanism driving superrotation in Titan's atmosphere \citep{Wang2014}.

In our simulations, there are two distinct types of
superrotation. During spin-up, we get superrotation at the equator
at the pressure levels of radiative heating, which is likely due to
the stationary wave response to the radiative forcing. Since we are
strongly in the weak temperature gradient regime (with a low mean
molecular weight atmosphere and slow rotation rate), the Rossby radius of
deformation is large, and there is only a weak excitation of equatorial
Rossby and Kelvin waves required to drive superrotating jets, which
are often seen in simulations of tidally-locked terrestrial exoplanets
\citep[e.g.][]{Pierrehumbert2019}. This is most likely the reason we
don't see an on-equator maximum of zonal wind in the lower atmosphere
(at around $10^4-10^5$ Pa).

\begin{figure*}
  \plotone{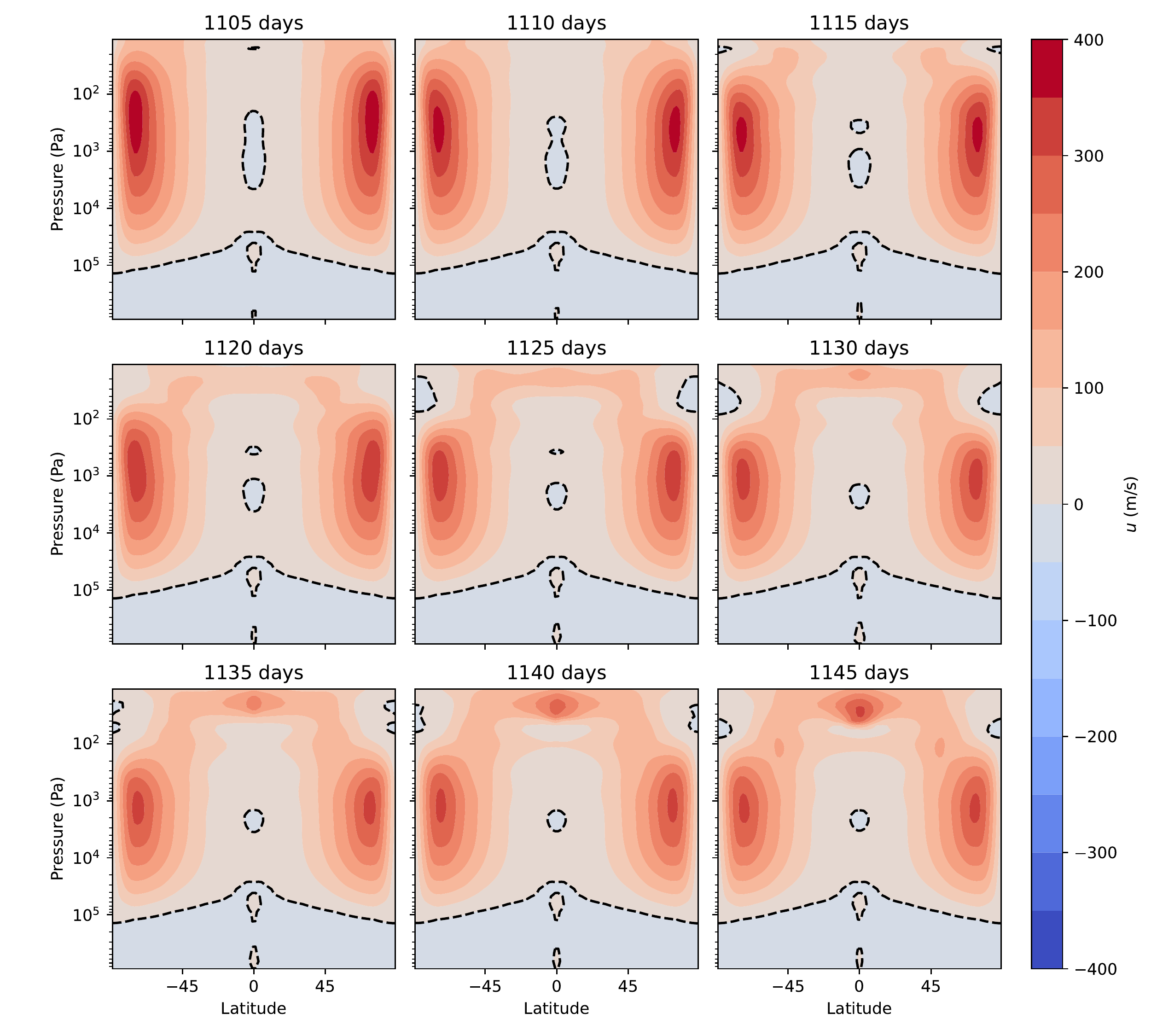}
  \caption{The zonal-mean zonal wind of the P6h run during the
    development of the transient instability causing
    superrotation. On the timescale of tens of days, momentum in the
    upper atmosphere is transferred from the poles to the
    equator. The $u=0$ contour is marked in dashed black.} \label{fig:time_inst}
\end{figure*}

In the upper atmosphere, however, we see a sudden transition to
superrotation caused by an instability. Figure~\ref{fig:time_inst}
shows the development of superrotation in the upper atmosphere over a
period of $\approx 40$ days, where momentum is transported from the
high latitude jet towards the equator. 

We performed additional experiments to test whether the location of
the upper boundary or the damping in the sponge layers affected the
presence of the instability. The model top of a GCM provides a
challenge, since often non-physical boundary conditions have to be
specified to ease solving the governing equations numerically. In ExoFMS,
there is an $\omega = 0$ boundary condition that keeps the model top
fixed at a constant pressure. This can lead to the spurious reflection
of waves from the model top which are removed by introducing some form
of damping
to the top model layers (known as a sponge layer). Our model gradually increases the divergence
damping coefficient towards the model top as:
\begin{equation}
  \label{eq:2}
  d_2 = d_{2,0}\qty[1-3\tanh(\frac{1}{10}\log(\frac{p}{p_0}))]
\end{equation}
where $d_{2,0}$ is a constant coefficient, set to 0.02 in our
simulations -- large enough to maintain model stability but small
enough not to affect the angular momentum budget of the atmosphere \citep{Lee2021}. In
the top two layers, the value of $d_{2,0}$ is doubled, then quadrupled
respectively to provide a stronger sponge layer damping.

To test the robustness of the superrotation, firstly we reduced the
model-top pressure by a factor of 100, from \SI{10}{Pa} to
\SI{0.1}{Pa}. We found that the instability still occurred, but at a
lower pressure of around \SI{2}{Pa} (c.f. \SI{40}{Pa} in the 10 Pa
model top simulations). In the 33 day period experiments, we noted
that the onset of superrotation occurred at a much later time, usually
around 10000-13000 days, around the time when the high-latitude
zonal jets have extended to the top of the atmosphere. This, along
with our experiment changing the model top pressure, suggests that
model-top effects could be playing a role in triggering the instability.

Secondly, we changed removed the pressure dependence of $d_{2,0}$ and
set it to its minimal value. This had little effect on the outcome of
the simulations, with the instability occurring at a similar pressure
level and time as the control. This proves that the instability
is not an artefact of the increase in divergence damping near the
sponge layer.

Lastly, we increased the depth and strength of the sponge layer, such
that the top five layers were sponge layers, and increased the damping
divergence coefficient to 0.05. In this experiment, the instability
still occurred, but the magnitude of the resulting equatorial zonal
wind was reduced by around a factor of two.

Note that divergence damping is only one choice of sponge-layer
mechanism. Another common choice is the presence of Rayleigh friction
in the top layers. However, we did not use this because it is known to
actively violate angular momentum conservation and have implications
for the upper atmosphere \citep{Shaw2007,Jablonowski2011}. We proceed with our analysis, noting that this feature of the
circulation seems to be robust to changes in damping but not ruling
out that the model top could affect the circulation.

\begin{figure*}
  \plotone{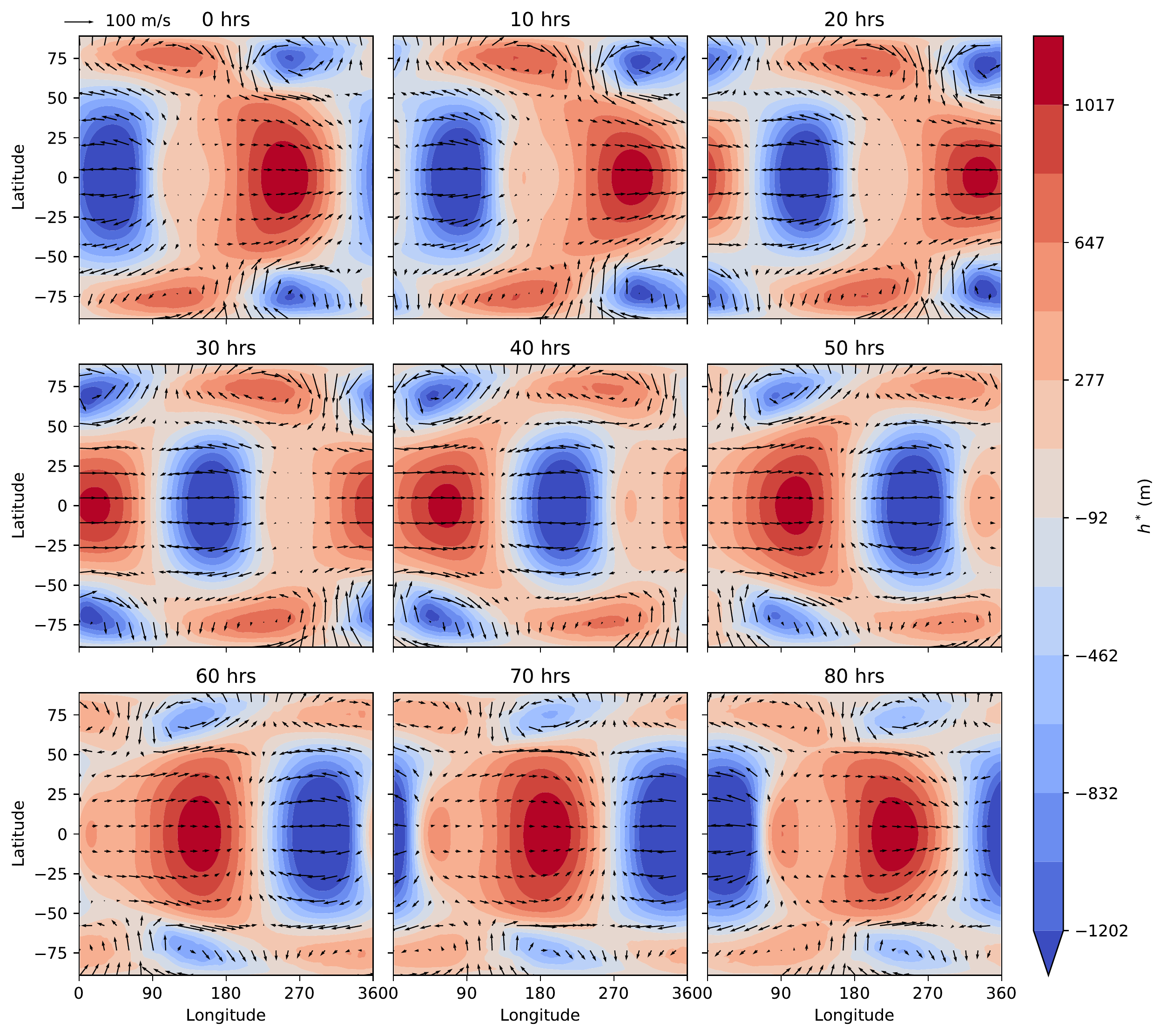}
  \caption{The zonal-eddy height field, $h^*$, across an 80 hour
    period of time during which the instability is forming. The arrows
  represent the zonal-eddy wind field. There is an eastward travelling
wave-like disturbance with zonal wavenumber 1. \label{fig:hstar}}
\end{figure*}

If we look at the zonal-eddy height field, $h^*$ as function of time
at the 40 Pa level (Figure~\ref{fig:hstar}),
we see that there is a eastward travelling disturbance which
couples the equatorial region to the vortices near the pole. We note
that the pattern of disturbance looks similar to the eddy height
fields seen in \cite{Mitchell2010} (Figure~14) and the pressure
perturbation fields in \cite{Wang2014} (Figures~1 and 2), suggesting
that a Rossby-Kelvin (RK) instability may be responsible for the
acceleration of the superrotation. The RK instability is caused by a
coupling of equatorial Kelvin waves and high-latitude Rossby waves,
and occurs in the regime where the Froude number of the flow (the
ratio between the frequency of the Rossby and Kelvin waves) is between
1-3 \citep{Wang2014}. The Froude number is given by:
\begin{equation}
  \label{eq:froude}
  \text{Fr} = \frac{U_0/\cos(\phi_{\text{max}})}{U_{\text{eq}} + NH/m},
\end{equation}
where $U_0$ is the high-latitude jet speed at latitude
$\phi_{\text{max}}$, $U_{\text{eq}}$ is the equatorial windspeed, $N$
is the buoyancy frequency, $H$ the scale-height of the disturbance and
$m$ the vertical wavenumber (i.e. $m=1$ represents the gravest
baroclinic mode of disturbance). Since the disturbance occurs in the
upper atmosphere of our model where the grey-gas equilibrium temperature
is constant, we can approximate $NH\approx \sqrt{RT_s}$
where $T_s$ is the skin temperature of the atmosphere. Neglecting the
equatorial wind speed during instability, we approximate Fr using
$U_0\approx$ \SI{400}{ms^{-1}}, $\phi_{\text{max}}\approx 75^{\circ}$
and $T_s \approx$ \SI{230}{K} which gives a Froude number of $\approx
1.6$, placing the atmosphere in the correct parameter regime for the
RK instability to take place.

\begin{figure*}
  \plotone{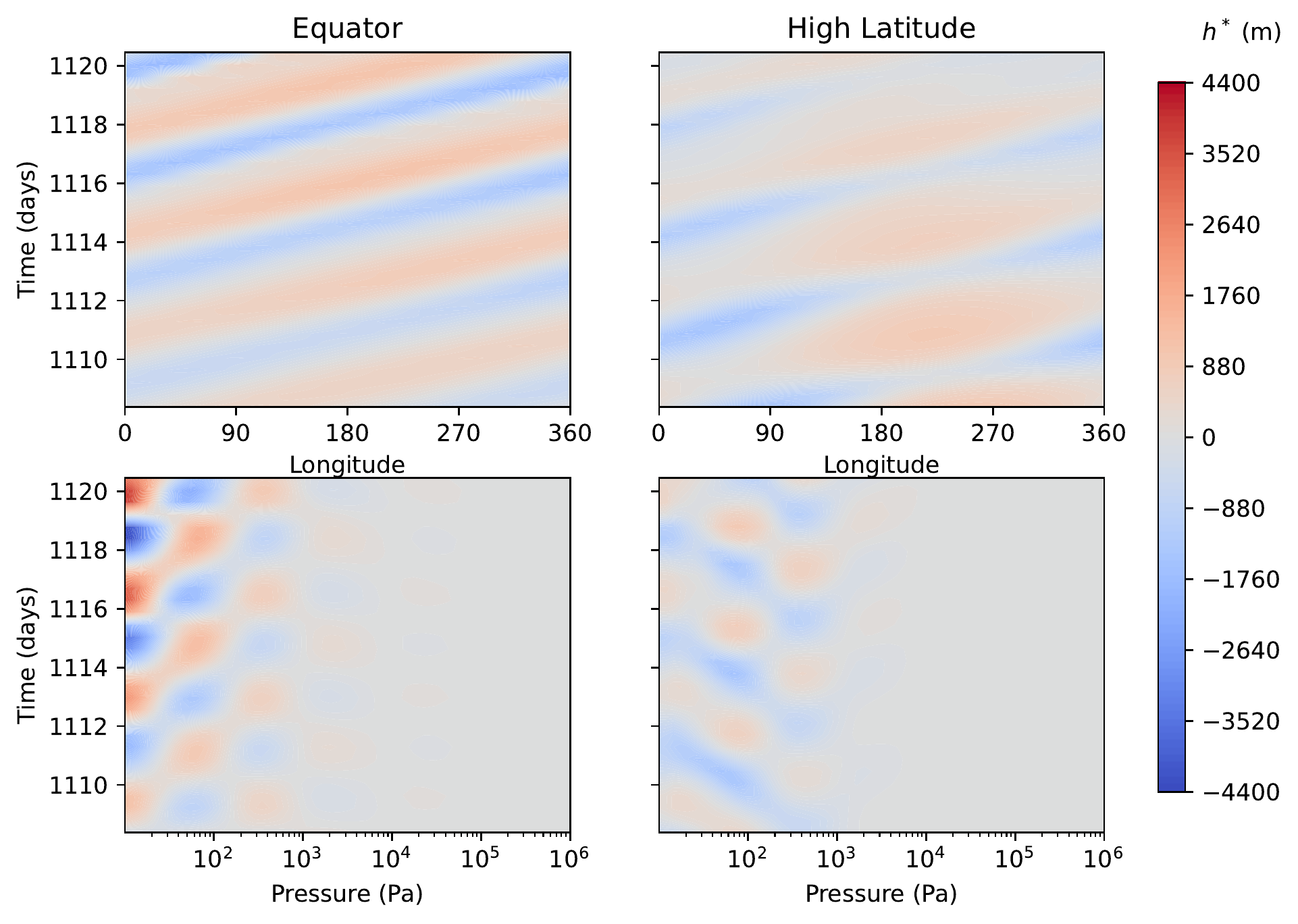}
  \caption{Hovm\"{o}ller diagrams of the zonal-eddy height field,
    $h^{*}$, averaged over the equatorial region (left column) and
    high-latitude regions (i.e. 50-80 degrees, right column). Top row:
    There is a clear zonal wavenumber 1 wave in the height field
    propagating eastwards. This structure is also clear at high
    latitudes, however there is also some beating in the response. Bottom row: The vertical structure of the waves
    at the equator suggests that the waves may be standing in the
    vertical direction, however the vertical structure at
    high-latitudes is more complicated. \label{fig:hovmoller}}
\end{figure*}

Figure~\ref{fig:hovmoller} shows the Hovm\"{o}ller diagram of the
zonal-eddy height field. There is a clear Eastwards travelling wave
pattern in the horizontal equatorial structure with constant phase
velocity. In the
high-latitude regions, this structure remains but with an added
beating in the pattern. Performing a Fourier transform in both time
and space, we extract the phase speed of the eastward travelling
component, and find it to be around \SI{330}{ms^{-1}} at the equator, comparable to the
speed of the high-latitude jets and to a characteristic Kelvin-wave
timescale, which is on the order of hundreds of
\si{ms^{-1}}. Performing this Fourier transform on the high-latitude
data also confirms that the beating pattern in the high-latitude data
is at 0 frequency, and represents the zonal wavenumber 1 time-mean
background in the height field. The vertical structure of the wave-pattern is
more complicated, but somewhat resembles a standing wave structure
near the equator, suggesting the instability could be equivalent-barotropic in nature.

\begin{figure}
  \plotone{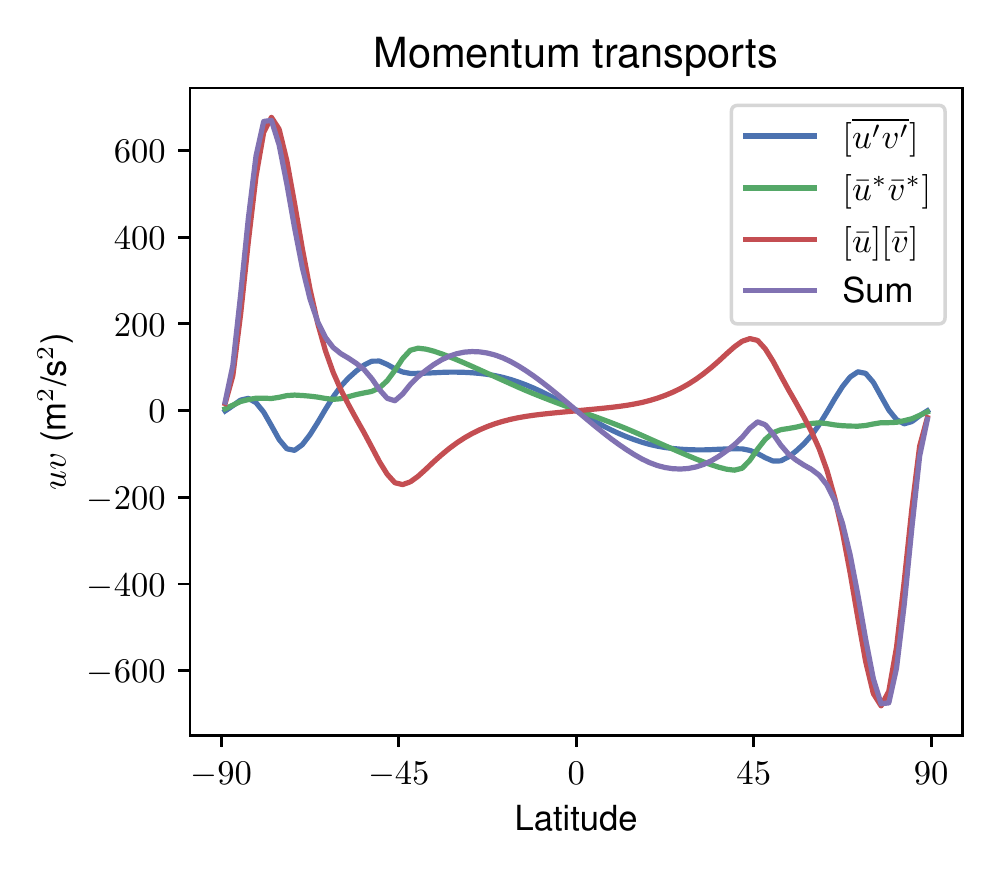}
  \caption{Terms in equation~\ref{eq:eddyeq} at the 40 Pa level. A
    negative gradient at the equator represents a convergence of
    eastwards momentum, causing an acceleration of the mean flow. \label{fig:fluxes}}
\end{figure}

Lastly, we show that this wave structure directly leads to the
acceleration of zonal wind at the momentum. If we decompose the
zonal-mean zonal wind equation into mean, zonal-eddy and time-eddy
components, we get:
\begin{widetext}
\begin{equation}
  \label{eq:eddyeq}
  \pdv{[\bar{u}]}{t} =
  -\frac{1}{a\cos^2\phi}\partial_\phi\Big\{\big(\underbrace{[\bar{u}][\bar{v}]}_{\substack{\text{mean}
    \\ \text{flow}}}
  + \underbrace{[\bar{u}^*\bar{v}^*]}_{\substack{\text{stationary}\\ \text{eddies}}} 
  + \underbrace{[\overline{u'v'}]}_{\substack{\text{transient}\\ \text{eddies}}}\big)\cos^2\phi \Big\}
  + \omega~\text{terms, Coriolis terms etc.}
\end{equation}
\end{widetext}
where overlined and bracketed terms represent time and zonal means
respectively, and starred and dashed terms represent deviations from
time and zonal means respectively. This equation represents the
meridional flux of momentum due to the mean flow, stationary eddies
and transient eddies on the mean jet speed. Figure~\ref{fig:fluxes}
shows the three horizontal terms in equation~\ref{eq:eddyeq} as a
function of latitude. The mean flow acts to transport
momentum away from the equator (positive gradient), whereas both the
stationary and transient eddies pump momentum towards the equator,
with the eddy accelerations dominating helping to accelerate the
equatorial jet at this level.

\section{Discussion} \label{sec:discussion}
\subsection{Comparision to other studies}
\label{subsec:comp}

On a qualitative level, our results look most similar to previous
studies of sub-Neptune GJ-1214b with solar metallicity, such as those
found in \cite{Kataria2014} and \cite{Drummond2018}, who find
high-latitude jets and also superrotation near the top of the
atmosphere. Our results also look qualitatively similar to those in
\cite{Wang2020} at earlier model times (around 1000-2000 days), where
they find two distinct high-latitude jets. However, in \cite{Wang2020} the
jets collapse into one wide equatorial jet as the kinetic energy
converges to equilibrium, which is not seen in our model. We note that
although GJ-1214b is similar in radius to K2-18b, it has a rotation
period of \SI{1.58}{days} \citep{Berta2011} so we would expect much
more pronounced equatorial dynamics and a stronger stationary wave
acceleration of the equatorial jet. The greater instellation (\SI{23600}{Wm^{-2}}
\citep{Wang2020}) has a much smaller effect on the WTG parameter
(which goes as $T^{1/2}\sim S_0^{1/8}$) but will affect the dynamics
via the radiative timescale, which will be much shorter and therefore
pronounce day-night temperature contrasts
\citep{Zhang2017}. Estimating the WTG parameter, $\Lambda= c_0/(\Omega
a)\approx \sqrt{RT}/(\Omega a)$ (the last assumption being valid if we
assume the dynamics occurs on a height scale greater than the
atmospheric scale height) for GJ1214 b using values found in \cite{Wang2020}, 
we find $\Lambda = 1.8$ for solar metallicity. Thus under the
assumption of a low mean molecular weight atmosphere, we might expect
to see some qualitative features of the WTG regime in this atmosphere.

In studies of tidally-locked terrestrial planets, the regime in which
the global Rossby radius exceeds planetary radius (i.e. the
``slow-rotator'' regime in \cite{Haqq-Misra2018} and ``Type I''
circulation in \cite{Noda2017}) is associated with the predominant
circulation being directly from dayside to nightside and isotropic
around the substellar point. This contrasts with our finding of strong
high-latitude jets and a residual equator-to-pole overturning circulation. However, we
note that our model is optically
thick to shortwave radiation, whereas terrestrial planet atmospheres are often
optically thin to SW radiation and have a significant proportion of
incoming stellar radiation absorbed at the surface. This drives
convection and convectional heating of the troposphere, which may
explain some disparities between models. Secondly, the surface is
often parametrised with some form of drag at the bottom boundary,
which isn't present in our model. Lastly, terrestrial planets are
typically modelled with an \ce{N2} dominated atmosphere. Increasing
the molecular weight of the atmosphere decreases $\Lambda$ (since
$R\propto \mu^{-1}$) and also decreases the heat capacity (which is
proportional to $R$ in the case of an ideal gas with a constant number
of degrees of freedom) which in turn shortens the radiative timescale
and can lead to higher day-night temperature contrasts.

The only other GCM study of K2-18b \citep{Charnay2021} (from now on referred to as C21), found a weak equatorial jet
(speed around \SI{50}{m/s}) at the 0.1-1 bar level and no high-latitude prograde jets for
their solar metallicity runs (which are most comparable to the PKc
experiment in this paper). The atmosphere was dominated
by the overturning circulation from day to night. This is in marked
contrast to our simulations, where the strongest jets were
high-latitude and the overturning day-night circulation, although
being the dominant overturning circulation, was small in comparison to
the rotational circulation. We find that the magnitude
of equatorial jet speed at the level of heating (around 0.1 to 1 bar)
in our PKc run ($\approx$ \SI{40}{m/s}) to be similar to the
$\approx$ \SI{50}{m/s} strength equatorial jet in C21. A significant difference in the model of
C21 is their inclusion of real-gas radiation, in
comparison to our grey-gas scheme. This leads to a cooler troposphere
in their model, along with larger vertical temperature gradients. One
way this could influence the dynamics is in the setting of the Rossby radius, stated as:
\begin{equation}
  \label{eq:4}
  L_R = \sqrt{\frac{NH}{2\Omega}}
\end{equation}
in C21, where $N\equiv \sqrt{g/T(g/c_p + \dv*{T}{z})}$
is the buoyancy frequency and $H$ the atmospheric scale height. We
find that between 0.1-1 bars, C21 has vertical temperature gradients
closer to adiabatic, leading to our value of $N$ being approximately 3
times larger than theirs, with relatively good agreement higher in the
atmosphere. With a smaller $L_R$, Rossby waves are trapped closer to
the equatorial region and we would expect a stronger stationary wave
response, and perhaps a stronger jet. However, given the qualitative
similarity between past studies using grey-gas and real-gas radiation
\citep[e.g.][]{Lee2021, Komacek2019b, Kaspi2015}, it would be
surprising if the different radiation schemes accounted for all of the
difference between the models. Other factors that could explain the
disparity in wind structure include the type and strength of numerical
damping schemes used and the type of dynamical core. ExoFMS uses a
cubed-sphere grid, whereas the LMD generic GCM used in C21 uses a
latitude-longitude grid, which could produce variation between the models in the
polar region where the latitude-longitude grid contains a
singularity. The presence of high-latitude jets should be a robust
result of transporting angular momentum from equator to poles, and our
model should in theory produce less distortion at the poles due to the
lack of coordinate singularity. 

\begin{figure*}
  \plotone{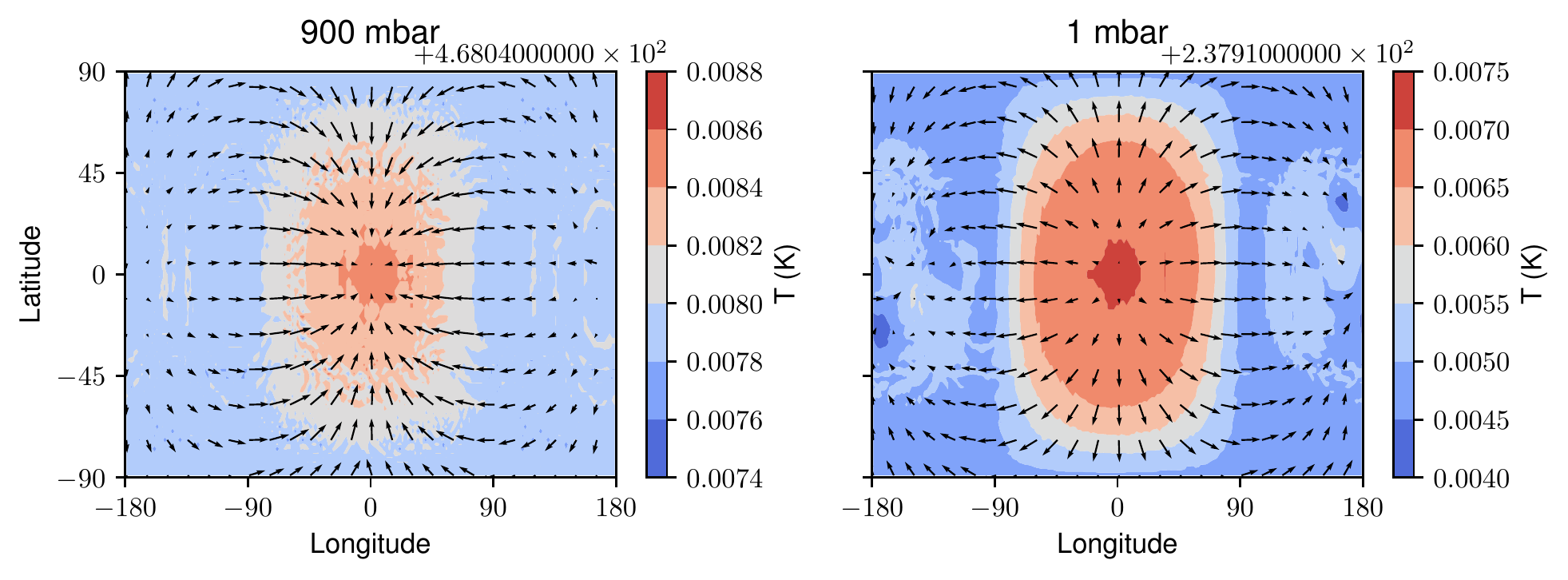}
  \caption{The temperature (contours) and wind fields (arrows) for the
  non-rotating model run. Left (1 bar level): Isotropic convergence of
wind at the substellar point and divergence at the antistellar
point. Right (0.01 bar level): In the upper atmosphere the winds
diverge around the substellar point and converge at the antistellar
point. \label{fig:no_rot}}
\end{figure*}

To test whether the poles are causing any distortion in the model, we
ran the model with no rotation, which should in theory produce
circulation symmetric around the substellar
point. Figure~\ref{fig:no_rot} shows the temperature and wind fields
in the lower and upper atmosphere, which exhibit the expected
isotropic divergent flow around the substellar and antistellar point.

\subsection{Validity of the Primitive Equations}\label{subsec:prim}
In \cite{Mayne2019}, the validity of using the primitive
  equations for GCM simulations of hot super-Earths and sub-Neptunes
  was investigated. Simulations using the full Navier-Stokes equation in which $Z/a\ge 20 \%$ (where $Z$ is
  the height of the dynamically active part of atmosphere and $a$ the
  planetary radius) begin to
  show significant deviations from the primitive equations due to the
  violation of the ``shallow approximation''. In our experiments, the
  maximum value of $Z/a$ was 8\% for the two highest-instellation
  cases, justifying the use of the shallow
  approximation. We might expect this given our atmospheres are cooler
  than those in \cite{Mayne2019}, with the scale height scaling with
  $T$. \cite{Mayne2019} also discuss the violation of the
  traditional approximation, where metric terms in the momentum
  equation become important. Similar to the method in
  \cite{Mayne2019}, to check if $v\tan\phi\gg w$ (i.e. the traditional
  approximation is valid), we looked at the sign of $v\tan\phi/(10w) -
  1$ at different pressure levels for each of our experiments. We
  found that $v\tan\phi\gg w$ in all experiments apart from a narrow
  narrow band at the equator (around $\pm 10^{\circ}$ latitude), where
$\tan\phi\ll 1$ and the traditional approximation is always
invalid. As discussed in \cite{Mayne2019}, if this region is limited
and connected smoothly to the mid-latitude flow, then the
approximation is acceptable. With weaker radiative forcing than
\cite{Mayne2019}, we would expect $w$ to be smaller and therefore the
traditional approximation is likely to be a good assumption in our
experiments. Nonetheless, further investigation of the potential
dynamical consequences of the non-traditional geometric terms for
sub-Neptune atmospheres would be a fruitful subject for future investigation.

\subsection{Scaling $\Lambda$ for general sub-Neptunes}
We now consider how the WTG parameter depends on stellar and planetary
parameters, to see if we can expect most temperate sub-Neptunes to
have $\Lambda \gg 1$. If we use the equilibrium temperature $T_{eq}\sim (S_0/\sigma)^{1/4}$
in estimating $\Lambda$, we note:
\begin{equation}
  \label{eq:stellar}
  S_0 = \frac{F}{4\pi d^2}
\end{equation}
where $F$ is the stellar luminosity and $d$ is the planet's orbital
distance. Using Kepler's third law (note for a tidally-locked planet
the orbital frequency is equal to the planet's rotation frequency):
\begin{equation}
  \label{eq:7}
  d^2 = \qty(\frac{GM_s}{\Omega^2})^{\frac{2}{3}},
\end{equation}
where $M_s$ is the stellar mass, we find:
\begin{equation}
  \label{eq:6}
  S_0 = \frac{F}{4\pi G^{\frac{2}{3}}} M_s^{-\frac{2}{3}}\Omega^{\frac{4}{3}}.
\end{equation}
If we are assuming $\Lambda = \sqrt{RT}/(\Omega a)$ this gives us an
approximate scaling:
\begin{equation}
  \label{eq:8}
  \Lambda \sim R^{\frac{1}{2}}F^{\frac{1}{8}}M_s^{-\frac{1}{12}}\Omega^{-\frac{5}{6}}a^{-1}
\end{equation}
We note that for M-type stars, $F$ varies from around
\num{3e-4}$L_\sun$ to 0.069 $L_\sun$
(making $(F_{\text{max}}/F_{\text{min}})^{1/8}\approx 2$), and $M_s$
varies from 0.08 to 0.57 $M_\sun$ (making
$(M_{s\text{,max}}/M_{s\text{,min}})^{1/12}\approx 1.2$)
\citep{Pecaut2013}. If we restrict our focus to sub-Neptunes, $a$
varies from around 2 to 4 \re, i.e. by a factor of two. Therefore, we expect that $\Lambda$ varies
mostly due to the global rotation rate $\Omega$, and the composition
$R\propto \mu^{-1}$, since $M_s^{-1/12}$ and $F^{1/8}$ do not vary
significantly across the range of M-dwarf stars and $a^{-1}$ does not
vary significantly for sub-Neptunes.

\begin{figure*}[t]
  \plotone{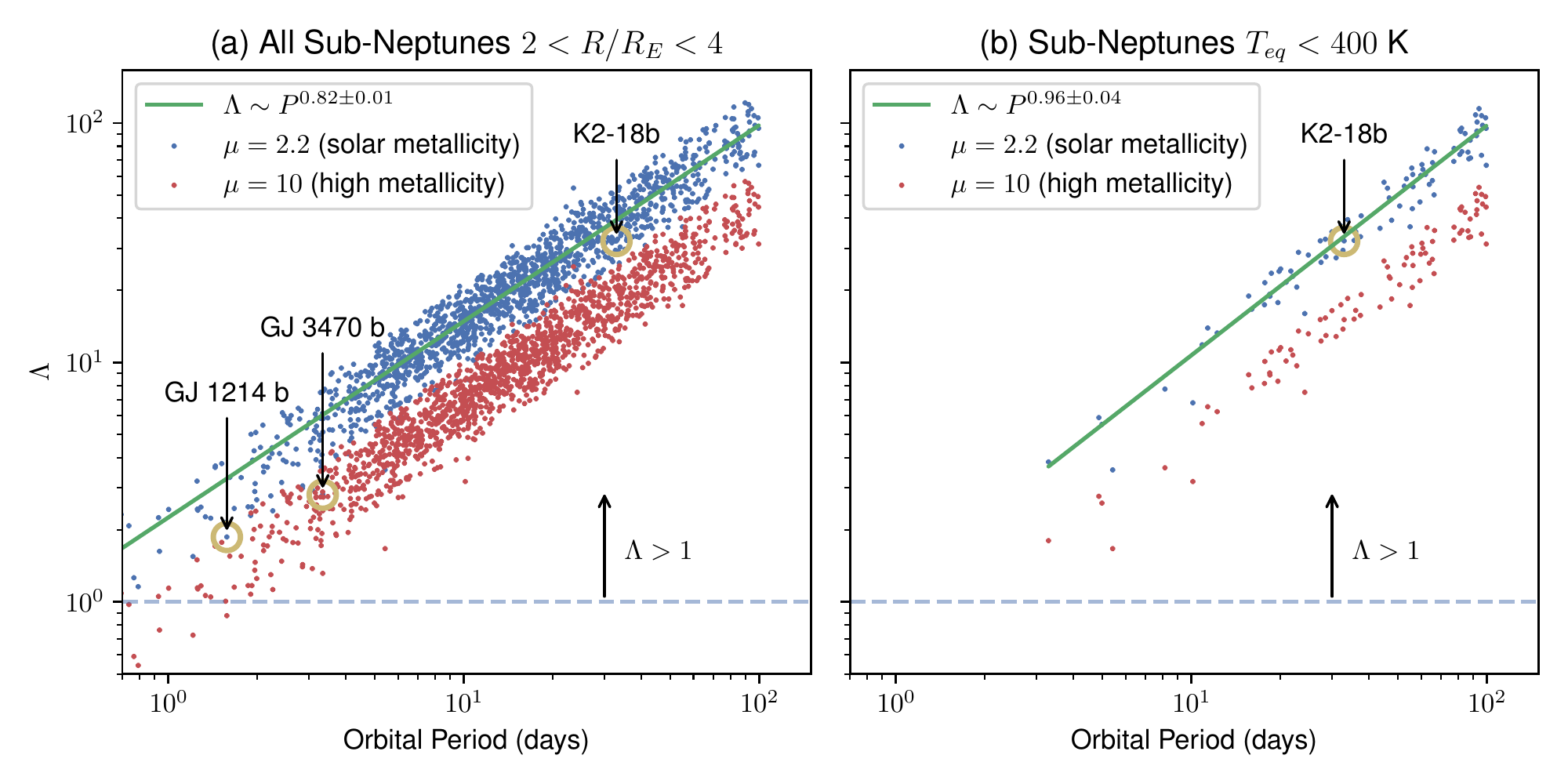}
  \caption{The weak temperature gradient parameter for observed
    sub-Neptunes with orbital period $<100$ days, assuming
    tidal-locking. The blue data series assumes a mean-molecular
    weight of 2.2 (c.f. this study), whereas the red series has
    $\mu=10$ to represent a higher metallicity atmosphere. (a) For all
    sub-Neptunes, we have $\Lambda>1$ in the majority of cases, and that the scaling
  $\Lambda\sim P^{0.83}\approx P^{5/6}$ is in agreement with
  equation~\ref{eq:8}, confirming that parameters $a$, $M_s^{1/12}$
  and $F^{1/8}$
  do not vary greatly over the range of sub-Neptunes. We also highlight
  the value of $\Lambda$ for three example sub-Neptunes (assuming
  $\mu=2.2$). (b) Same as (a), but filtered for $T_{eq}<400$ K. Here
  $\Lambda >1$ for all cases, but the scaling for $\Lambda$ is less
  accurate suggesting other parameters require scaling for. \label{fig:wtgparam}}
\end{figure*}

Figure~\ref{fig:wtgparam}(a) shows $\Lambda$ for all discovered
sub-Neptune exoplanets with orbital period less than 100 days,
assuming tidal-locking\footnote{Data from the Nasa Exoplanet Archive,
  \url{https://exoplanetarchive.ipac.caltech.edu}, accessed
  \formatdate{10}{9}{2021}}. We plot $\Lambda$ for low-metallicity
($\mu=2.2$) and high-metallicity ($\mu=10$) atmospheres, and find that
in both cases $\Lambda \gg 1$ for the majority of sub-Neptunes. The
best fit power law to the data also shows $\Lambda\sim P^{0.83\pm
  0.01}$, which is in agreement with equation~\ref{eq:8}, and
confirms the analysis that other parameters such as stellar mass and
flux do not vary greatly over the range of sub-Neptune exoplanets.

We stress that having $\Lambda>1$ does not imply that all the planets
in Figure~\ref{fig:wtgparam}(a) will exhibit weak horizontal temperature
gradients. As discussed in \cite{Komacek2016}, \cite{Zhang2017} and
section~\ref{subsec:tempstructure}, temperature gradients also depend
on the radiative timescale. Since $\tau_r\propto T^{-3}p(\tau=1)$, planets with
high equilibrium temperatures can have very short radiative timescales
and will therefore exhibit strong day-night temperature contrasts. In
addition, high-metallicity atmospheres where the level of radiative
cooling, $p(\tau=1)$, is high in the atmosphere due to increased
longwave opacity, will also have short
radiative timescales. In Figure~\ref{fig:wtgparam}(b), we plot
sub-Neptunes for which the equilibrium temperature, $T_{eq}$, is less
than 400 K. Here $\Lambda>1$ always, and since $T$ is lower we can be
more confident these sub-Neptune atmospheres will have long radiative timescales
similar to those in this study and exhibit similar dynamics to our
simulated atmospheres.

\subsection{1D vs 3D Modelling}\label{subsec:1dvs3d}

The magnitude of the horizontal temperature gradients provides a
useful bound on the error of 1D column models of sub-Neptune
atmospheres. For applications that do not depend on the structure of
the 3D circulation (which is not horizontally uniform), 1D models may
be able to give us useful insights at much lower computational
expense than 3D GCMs. For our 33 day period experiments, the maximum fractional
horizontal temperature difference from the substellar point
(i.e. $(T-T_{\text{substellar}})/T_{\text{substellar}}$) is around 6
to 8\%, rising to 15 to 22\% in the 6 day period experiments. Whether
this error is tolerable is context-dependent. For example, in phase
curve calculations the projected area of the polar regions (where the
temperature deviation from pure heat redistribution is greatest) is
small and deviations may not affect calculations significantly. On the
other hand, polar regions of the atmosphere would be important for
transmission spectroscopy measurements.

Understanding the 3D structure of these atmospheres is
crucial for some contexts. For example, estimates of the
``$K_{zz}$'' parameter in 1D chemistry models often rely on having
information on the characteristic vertical velocities in an atmosphere
\citep{Zhang2018}. Since the vertical velocities in a GCM are
controlled by the divergent circulation, scalings like
equation~\ref{eq:psiscale} which link the divergent circulation to
planetary parameters could be useful in estimating such
quantities. Moreover, transport processes important in moving around
chemical tracers and cloud particles are inherently 3D processes and
so may not be amenable to simple 1D modelling. For example,
  C21 similarly found weak horizontal temperature gradients. However,
  the inhomogeneous circulation can cause spatially and temporally
  varying cloud profiles. Though we might expect the relative
  horizontal uniformity of temperature to survive inhomogeneous cloud
  effects, because it derives from the basic WTG scaling, cloud
  feedbacks could nonetheless have an important influence on the
  vertical structure of the atmosphere.

\subsection{Model Limitations and Future Work}

The grey gas radiation used in our model is a crude approximation of
the full radiative transfer calculation. As discussed in the
introduction, we underestimate the temperature of the deep atmosphere
by around 100 K compared to real-gas 1D simulations. Moreover, grey
gas atmospheres have a tendency to be sub-adiabatic throughout the
atmosphere, underestimating the value of the vertical temperature
lapse-rate. This is the same effect that led to our overestimation of
$N$ compared to \citep{Charnay2021}, though we note that in C21 the
temperature profile is close to adiabatic in a very limited section of
the atmosphere. However, the grey-gas scheme produces qualitatively
similar temperature profiles to the real-gas models and should be
valid for looking at global-scale circulations. Moreover, the grey-gas
scheme is much less prohibitive in terms of its computational speed
compared to using real-gas radiation, and allows us to run several
comparative models of sub-Neptune atmospheres out to tens of thousands
of days.

In this study we have examined the dry dynamics of the atmosphere to
highlight the basic dynamical behaviour. The tentative detection of
water vapour in the atmosphere of K2-18b \citep{Benneke2019} and
subsequent modelling \citep{Blain2021,Charnay2021} have shown that
there are circumstances under which water can condense in the
atmospheres of sub-Neptunes. We will investigate the effect of
condensation on the dynamics in Part II.

We would also encourage an inter-comparison similar to the TRAPPIST
Habitable Atmosphere Intercomparison (THAI) project
\citep{Fauchez2021} to look at the differences between GCMs when
modelling a fiducial sub-Neptune exoplanet. The difference between
seemingly similarly set-up GCM experiments (e.g. between ours and
\citep{Charnay2021}) suggests that further collaboration is required
to illuminate the differences caused by different dynamical cores,
physical parametrisations and damping schemes. A good starting point would
be an intercomparison between grey models, to highlight the dynamics.

Lastly, more work could be done on the instability found in the upper
atmosphere to fully diagnose its origin and effect on the atmosphere,
perhaps in the form of simplified models or instability analysis \citep[e.g.][]{Wang2014}.

\section{Concluding Remarks}

In this work we have presented a suite of GCM simulations aimed at
modelling temperate sub-Neptune planets, using K2-18b as our control
experiment. Overall these atmospheres exhibit weak horizontal
temperature gradients. We found that these atmospheres are dominated by
high-latitude cyclostrophic jets, with weak equatorial superrotation
in the lower atmosphere and strong, instability-driven equatorial jets
in the upper atmosphere. We confirmed the result of \cite{Wang2020},
finding that our models have convergence times on the order of tens of
thousands of days, increasing with slower
rotation rate and lower instellation. We found that the high-latitude
jets are cyclostrophically balanced in our fast-rotating experiment,
with non-linear advective terms providing a significant proportion of
the balance in the slower-rotating, hotter experiments. Using the
framework of \cite{Hammond2021}, we decomposed the circulation into
divergent and rotational components in tidally-locked coordinates, and
found that although the dominant flow was rotational, the overturning
circulation was dominated by a day-night circulation responsible for
redistributing the energy deposited by instellation from dayside to
nightside. We provided a scaling argument for how the strength of this
circulation varies with instellation, finding that the tidally-locked
streamfunction should scale as $S_0^{3/4}$.

In all our experiments, we observed equatorial superrotation in the
upper atmosphere of our model, driven by an instability similar in
structure to the one modelled in \cite{Wang2014} to explain
superrotation in slow, non-synchronously rotating planets such as
Venus and Titan. We showed that this instability occurred in the
correct parameter regime as in \cite{Wang2014}, and that it provided
eddy-momentum fluxes which could transport zonal momentum from the
high-latitude jets towards the equator.

Finally, we compared our results to the literature, finding our
results differ qualitatively to \cite{Charnay2021}, who model K2-18b
using the LMD Generic GCM but do not observe high-latitude cyclostrophic jets in their data. We offered some reasons why our model could
produce different findings, but ultimately we recommend some form of
intercomparison between GCM models to illuminate the effects of
different physical parametrisations and dynamical cores on the
atmospheres of sub-Neptunes. We noted the need to consider moist
effects and leave that to the contents of a future study.

\acknowledgements
We thank Neil Lewis and Mark Hammond for letting us use their
code to calculate Helmholtz decompositions and Neil Lewis, Shami Tsai
and Mark Hammond for useful discussions and comments on the
manuscript. We thank the anonymous reviewer for their feedback, which
improved this manuscript. This work was supported by
grants from the European Research Council (Advanced
grant EXOCONDENSE \#740963 to R.T.P.)

\software{\textsc{numpy} \citep{Harris2020}, \textsc{xarray}
  \citep{Hoyer2017}, \textsc{scipy} \citep{Virtanen2020},
  \textsc{matplotlib} \citep{Hunter2007}, \textsc{windspharm}
  \citep{Dawson2016}}

\appendix

\section{Effect of pressure dependent shortwave opacity}\label{app:a}
We present here a comparison of the P6c experiment (see
Table~\ref{tab:exp}) to an identical one run with the SW opacity
proportional to $p$ (hereafter referred to as the pressure-depedent
opacity (PDO) experiment). We changed the opacity such that the total SW
optical depth of the atmosphere remained the same.

\subsection{Temperature Profiles}\label{subsec:tprofs}
\begin{figure}[!htbp]
  \epsscale{0.7}
  \plotone{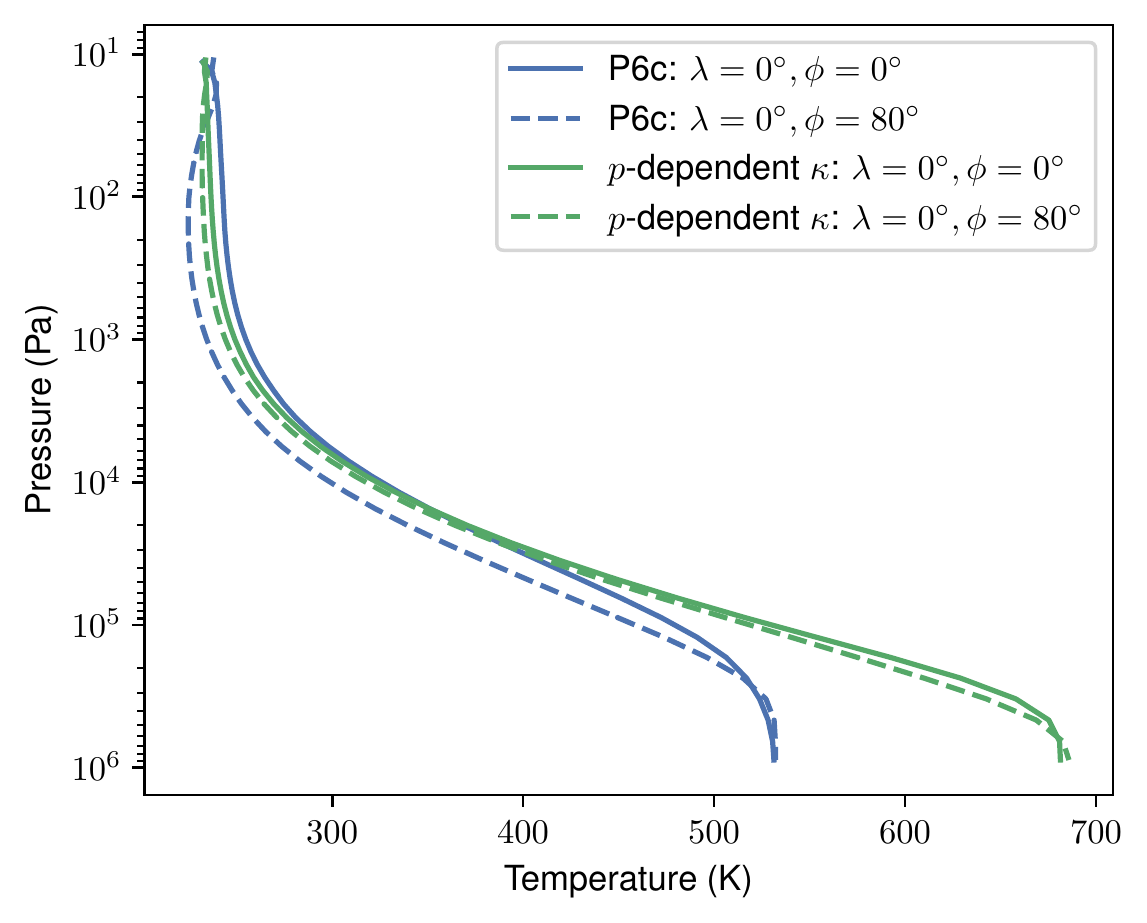}
  \caption{Latitudinal temperature variation of the two
    experiments. The case with pressure-dependent opacity shows a much
  smaller equator to pole temperature contrast.\label{fig:temp_pbroad}}
\end{figure}

As shown in Figure~\ref{fig:ggcomp}, including pressure-dependent SW
opacities increases the temperature at the bottom of the
atmosphere. It also increases the pressure of the characteristic SW
heating level where $\tau_{\text{SW}}=1$ by a factor of
$\sqrt{\tau_{\text{SW0}}}$, where $\tau_{\text{SW0}}$ is the total SW
optical depth of the atmosphere. In the context of our experiments,
this moves the $\tau=1$ level from 0.8 bar to 2.8 bar.
\medskip

In Figure~\ref{fig:temp_pbroad} we look at the latitudinal variation
in temperature in both experiments. We note that the longitudinal
variation at all latitudes is extremely small in both cases (on the
order of $10^{-2}$ K). We see that the drop in temperature between
equator and pole is smaller in the PDO experiment, which confirms that we are still well within the weak
temperature gradient regime discussed in the main text.

\subsection{Zonal Wind}\label{subsec:zwind_app}
\begin{figure}[!htbp]
  \plotone{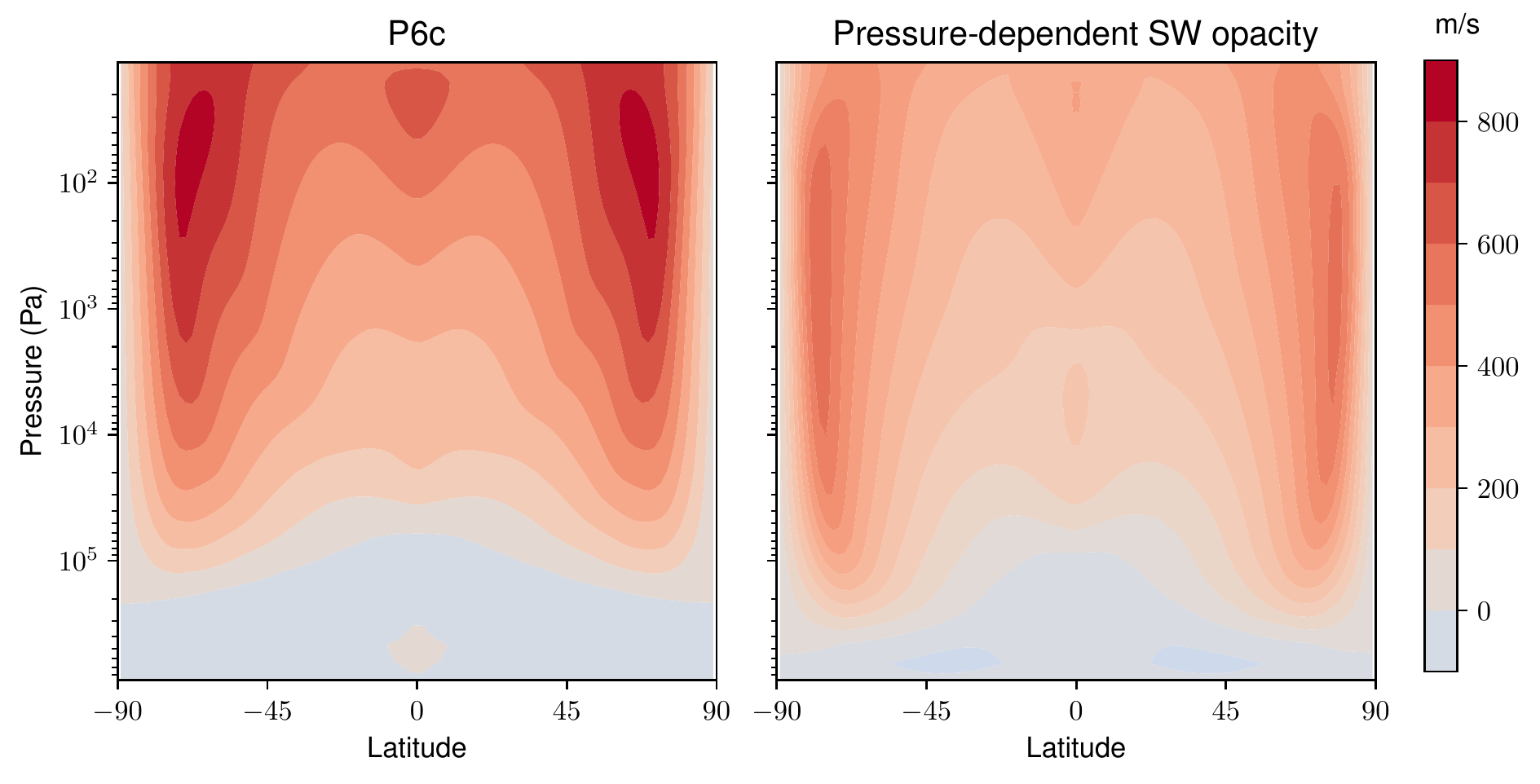}
  \caption{The zonal-mean zonal wind profiles.\label{fig:zwindpdf}}
\end{figure}

In Figure~\ref{fig:zwindpdf} we compare the zonal-mean zonal wind
profiles between the two experiments. The PDO experiment shows qualitatively similar zonal wind structure, with two
high latitude, cyclostrophically balanced zonal jets and equatorial
superrotation with a maximum in the upper atmosphere. We note that the
magnitude of the zonal wind in the PDO experiment is significantly lower than in the P6c experiment, which is
consistent with the lower equator-to-pole temperature difference seen
in section~\ref{subsec:tprofs}, since this is proportional to the
vertical wind shear in cyclostrophic balance. However, since we
currently have no theory predicting the strength of this wind and
temperature gradient a priori, we cannot explain why the magnitude of
this wind differs on changing the SW heating profile.
\medskip

We also note that the jets extend to higher pressures in the PDO
experiment. This makes sense since the characteristic level of SW
heating (discussed in section~\ref{subsec:tprofs}) is at a higher
pressure in this experiment, which drives the dynamics at this level.
\medskip

The instability discussed in section~\ref{subsec:superrotation} was
also present in the PDO experiment.

\subsection{Mass Streamfunctions}\label{subsec:mpsi_app}
\begin{figure}[!htbp]
  \plotone{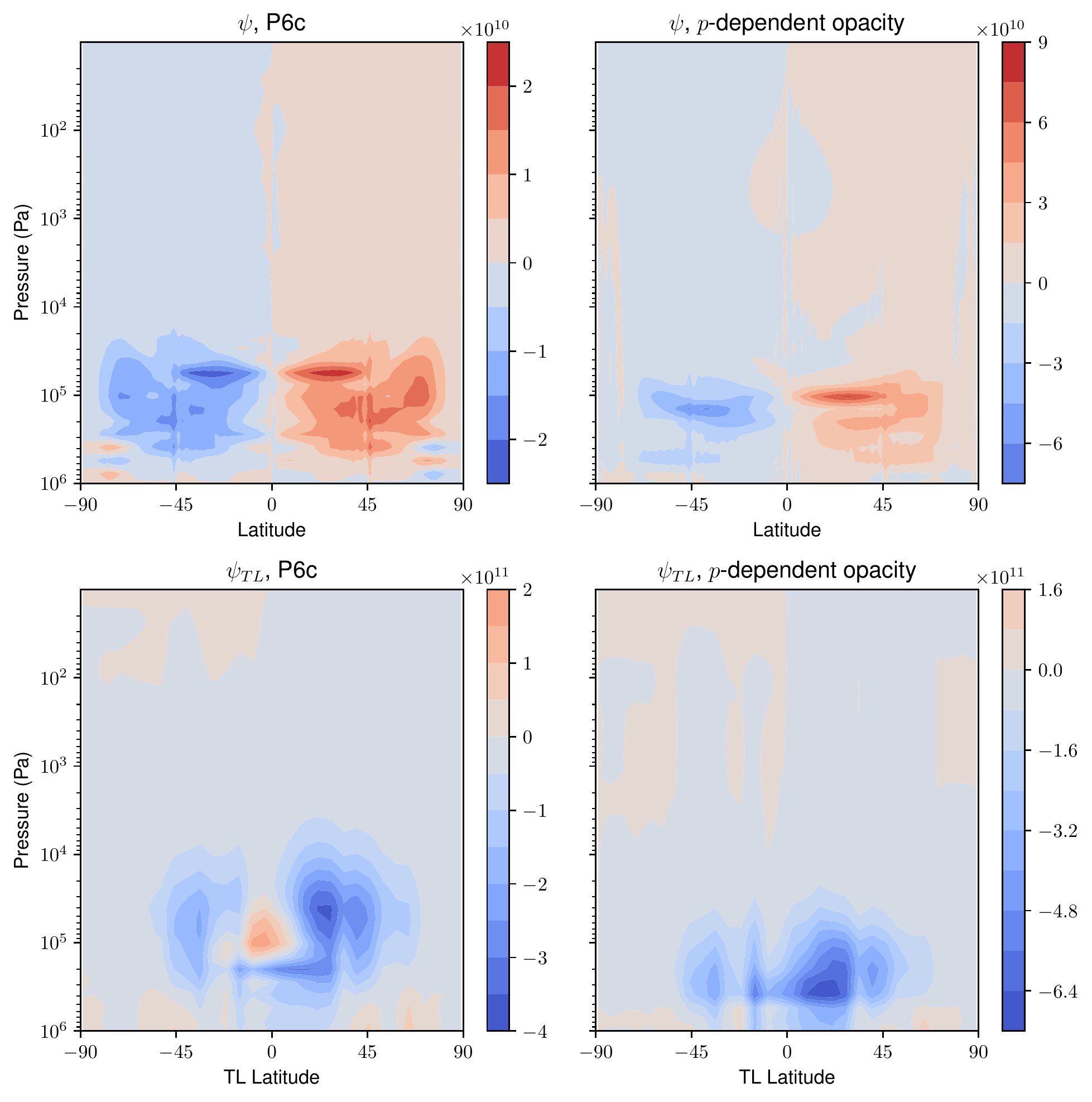}
  \caption{The mass streamfunction (in units of kg/s) in conventional latitude-longitude
    coordinates (top row) and tidally-locked coordinates (bottom row)
    for the P6c (left) and PDO (right).
    experiments. \label{fig:mpsi_pbroad}}
\end{figure}

Lastly, we compare the mass streamfunctions of the two
experiments. Qualitatively, the structure of the streamfunctions is
similar between the two experiments. In the PDO runs, the level of the
maximum streamfunction increases in pressure, which is linked once
again to the characteristic level of SW heating moving to higher pressures. The magnitude of the
tidally-locked streamfunction (which we related to the strength of the
instellation in section~\ref{subsec:overturning}) is of the same order
of magnitude but not identical between the two experiments. Our
scaling in section~\ref{subsec:overturning} linked the streamfunction
to the magnitude of the solar heating. Since the maximum magnitude of
the streamfunction is roughly at the level of characteristic SW
heating in both cases (0.8 bar for P6c, 2.2 bar for PDO), we would
expect the integrated heating at this level not to vary between
cases. However, changes in the mean dry static energy, and physics not
accounted for by the crude approximations made in deriving the scaling
law may cause the two values to be different.

\section{Temperature-pressure profiles}\label{app:b}

\begin{figure}[!htbp]
  \epsscale{0.9}
   \plotone{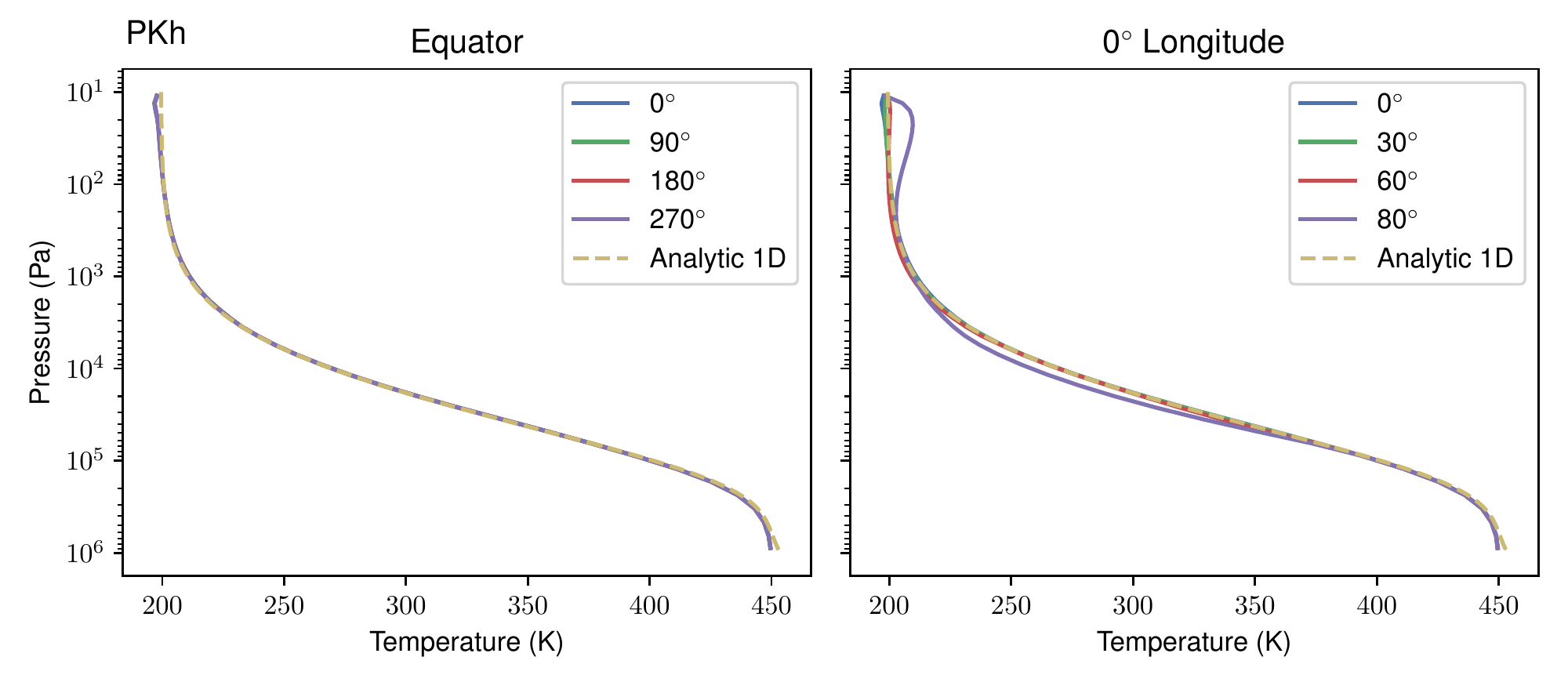}\\
   \plotone{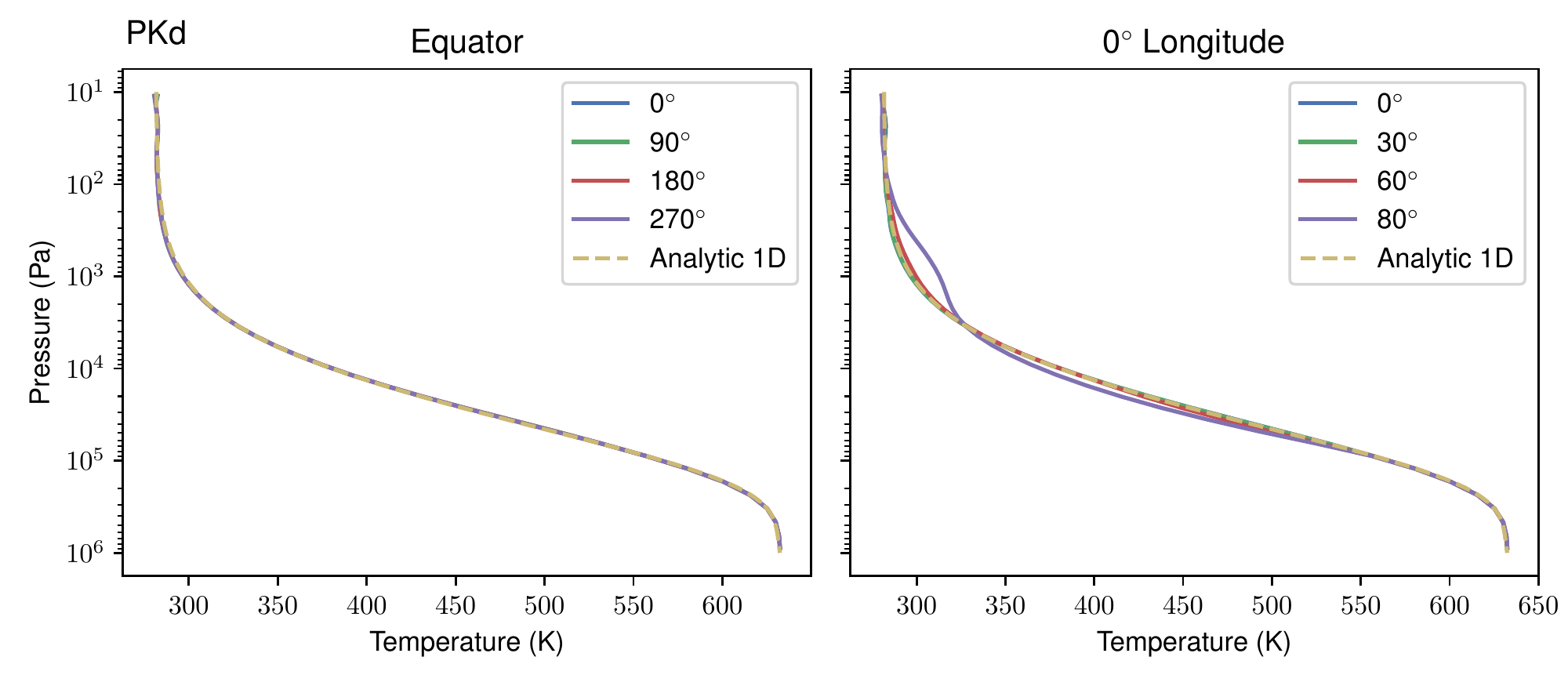}\\
   \caption{Temperature-Pressure profiles of the non-control, 33 day
     period experiments
     experiments. \label{fig:tempappend1}}
 \end{figure}

\begin{figure}[!htbp]
   \epsscale{0.9}
   \plotone{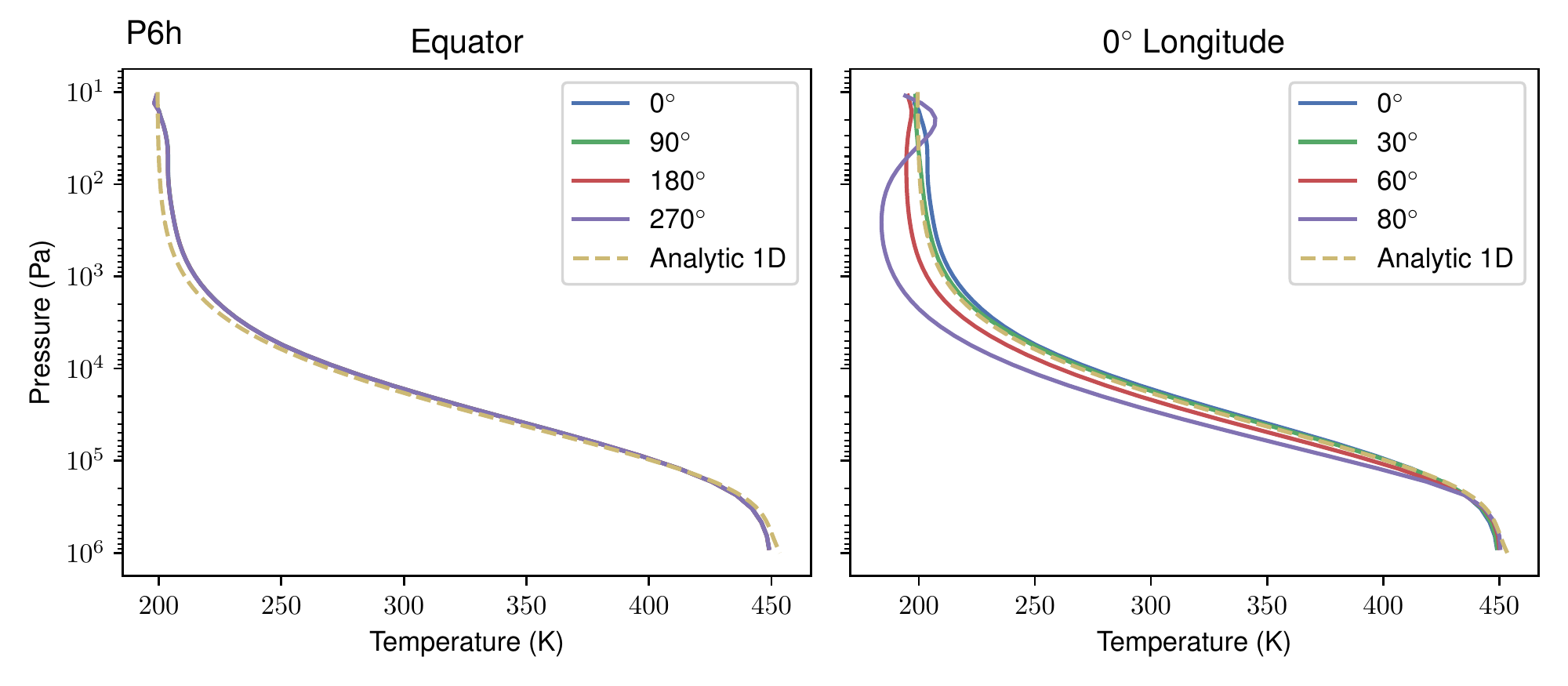}\\
   \plotone{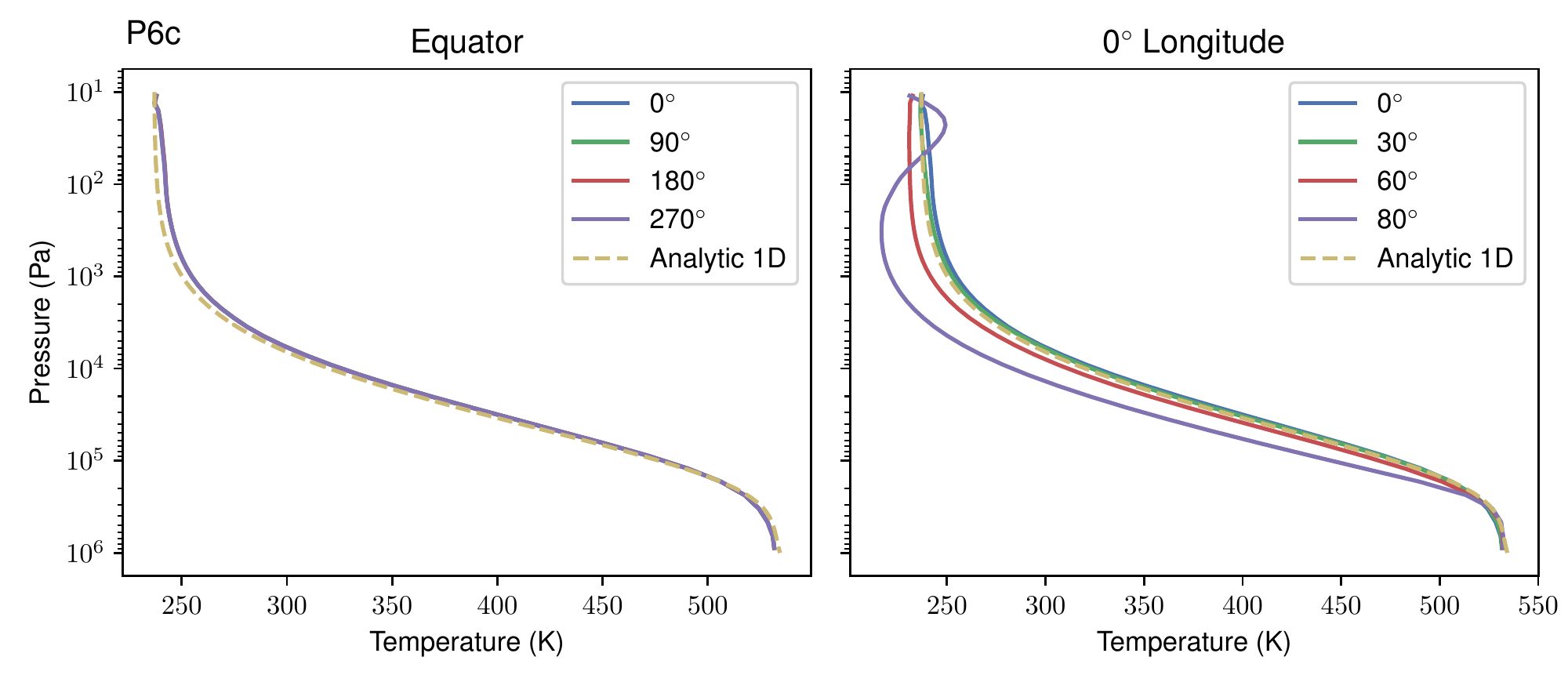}\\
   \plotone{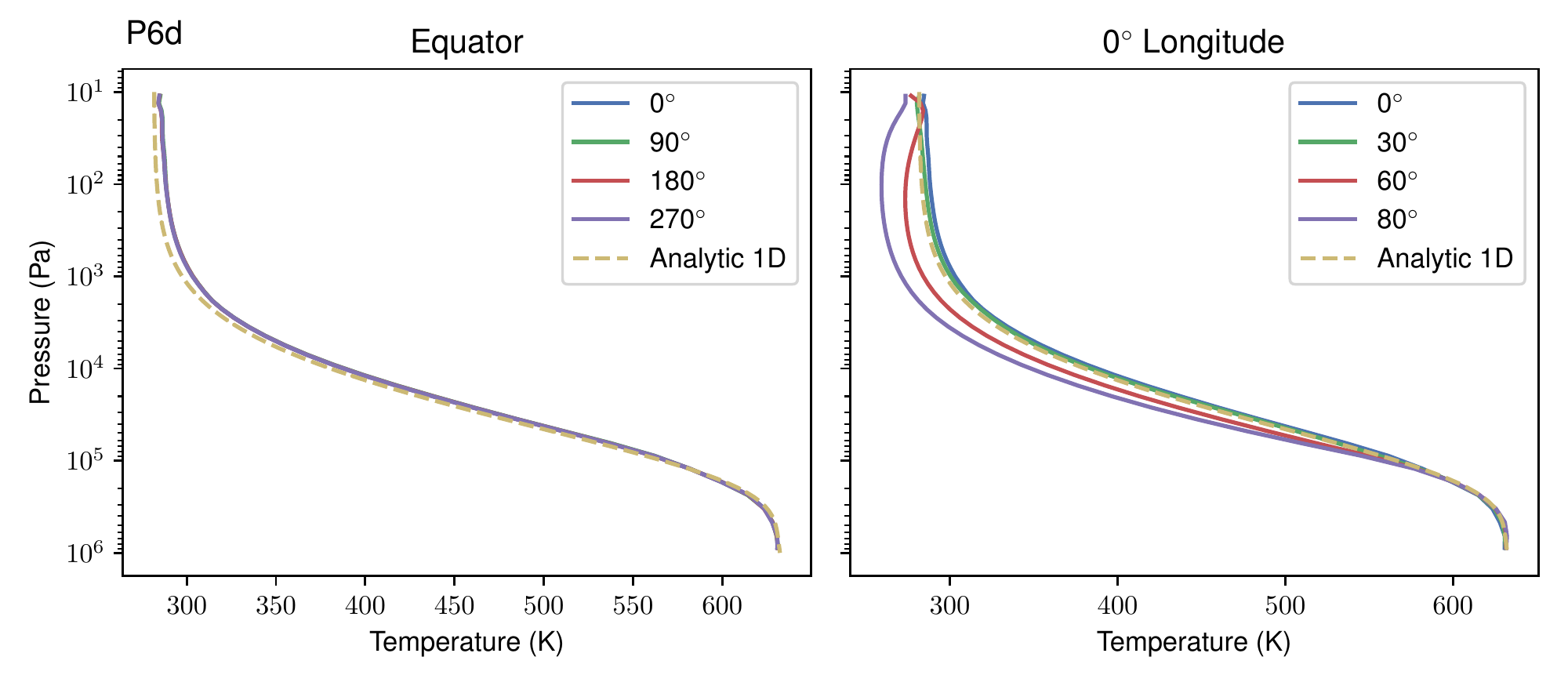}

   \caption{Temperature-Pressure profiles of the non-control, 6 day
     period experiments. \label{fig:tempappend2}}
 \end{figure}

We present in Figures~\ref{fig:tempappend1} and \ref{fig:tempappend2} the temperature-pressure profiles of the
non-control experiments (similar to Figure~\ref{fig:temp2}), for reference. 
 
 \clearpage
\bibliography{library,other}
\end{document}